\let\texyear\year
\documentclass{ieeeaccess}

\let\year\texyear
\usepackage{cite}
\usepackage{amsmath,amssymb,amsfonts}
\usepackage{algorithmic}
\usepackage{graphicx}
\usepackage{textcomp}

\usepackage{textcomp}
\usepackage{bm}
\makeatletter
\providecommand{\xfigwd}{\columnwidth}
\AtBeginDocument{\DeclareMathVersion{bold}
\SetSymbolFont{operators}{bold}{T1}{times}{b}{n}
\@ifundefined{symNewLetters}{}{\SetSymbolFont{NewLetters}{bold}{T1}{times}{b}{it}}
\SetMathAlphabet{\mathrm}{bold}{T1}{times}{b}{n}
\SetMathAlphabet{\mathit}{bold}{T1}{times}{b}{it}
\SetMathAlphabet{\mathbf}{bold}{T1}{times}{b}{n}
\SetMathAlphabet{\mathtt}{bold}{OT1}{pcr}{b}{n}
\SetSymbolFont{symbols}{bold}{OMS}{cmsy}{b}{n}
\renewcommand\boldmath{\@nomath\boldmath\mathversion{bold}}}
\makeatother

\def\BibTeX{{\rm B\kern-.05em{\sc i\kern-.025em b}\kern-.08em
    T\kern-.1667em\lower.7ex\hbox{E}\kern-.125emX}}

\usepackage{physics}
\usepackage{xcolor}
\usepackage{tabularx}
\usepackage{booktabs}
\usepackage{array}
\usepackage{hyperref}
\usepackage[edges]{forest}
\usepackage{svg}

\usepackage{tikz}
\usetikzlibrary{positioning, shadows}
\usetikzlibrary{shapes.geometric, arrows.meta}
\usetikzlibrary{shapes,arrows}
\usepackage{multirow, multicol}
\usepackage[footnote]{acronym}
\usepackage{fontawesome}

\NewSpotColorSpace{PANTONE}
\AddSpotColor{PANTONE} {PANTONE3015C} {PANTONE\SpotSpace 3015\SpotSpace C} {1 0.3 0 0.2}
\SetPageColorSpace{PANTONE}

\begin{document}
\history{Date of publication xxxx 00, 0000, date of current version xxxx 00, 0000.}
\doi{10.1109/TQE.2020.DOI}

\title{A Review on Quantum Satellite Communications: Challenges and Future Directions}
\author{\uppercase{Omar Alnaseri}\authorrefmark{1}, \IEEEmembership{Senior Member, IEEE},
\uppercase{Yassine Himeur}\authorrefmark{2}, \IEEEmembership{Senior Member, IEEE}, Ahmed Al Asadi\authorrefmark{3}, Mohammed A. Ala'anzy\authorrefmark{4}, \IEEEmembership{Senior Member, IEEE}, Rida Gadhafi\authorrefmark{2},~\IEEEmembership{Senior Member,~IEEE}, Shadi Atalla\authorrefmark{2},~\IEEEmembership{Senior Member,~IEEE}, and Wathiq Mansoor\authorrefmark{5},~\IEEEmembership{Senior Member,~IEEE}}

\address[1]{Department of Electrical Engineering, Baden-Wuerttemberg Cooperative State University, Ravensburg, Germany}
\address[2]{College of Engineering and Information Technology, University of Dubai, Dubai, United Arab Emirates}
\address[3]{Department of Communication Engineering, University of Technology- Iraq, Baghdad, Iraq}
\address[4]{Department of Computer Science, SDU University, Kaskelen 040900, Kazakhstan}
\address[5]{College of Engineering, American University of Iraq—Baghdad, Baghdad 10071, Iraq}
% \tfootnote{This paragraph of the first footnote will contain support
% information, including sponsor and financial support acknowledgment. For
% example, ``This work was supported in part by the U.S. Department of
% Commerce under Grant BS123456.''}

% \markboth
% {Author \headeretal: Preparation of Papers for IEEE TRANSACTIONS and JOURNALS}
% {Author \headeretal: Preparation of Papers for IEEE TRANSACTIONS and JOURNALS}

\corresp{Corresponding author: Omar Alnaseri (e-mail: alnaseri.omar@dozent.dhbw-ravensburg.de).}

\begin{abstract}
Quantum satellite communication (QSC) is emerging as a strategic technology for secure global networking and long-distance quantum connectivity. This review prioritizes the major challenges that still hinder large-scale deployment, including atmospheric loss, beam pointing and tracking, payload constraints, synchronization, scalability, and integration with terrestrial infrastructure. To contextualize these issues, we provide only a concise overview of the core concepts and enabling technologies behind QSC, together with representative milestones such as the Micius mission. Building on this background, the paper surveys recent advances in protocols, hybrid space--terrestrial architectures, turbulence mitigation, and AI-assisted optimization. It then examines future directions, including quantum Internet integration, daylight operation, satellite-supported repeaters, and space-based quantum computing. By centering the discussion on open technical bottlenecks and emerging research trajectories, this review aims to support researchers and engineers working toward practical and resilient QSC systems.
\end{abstract}


\titlepgskip=-15pt

\maketitle

\section{Introduction}
\PARstart{Q}{uantum} is set to play a crucial role in the development of the future quantum Internet, where a global network is designed to link quantum devices via satellite. QSC harnesses the unique properties of quantum physics, such as entanglement and superposition, to allow for exceptionally secure communication, most notably through the quantum key distribution (QKD). 

The primary purpose of this review is to provide a focused and up-to-date synthesis of the field with emphasis on the practical barriers and research opportunities that define QSC today. Rather than offering an extended tutorial on quantum theory, the paper uses a concise technical background to frame the core engineering issues, recent advances, and future directions most relevant to deployable quantum-enabled satellite networks. Specifically, we highlight the technological milestones that have shaped QSC, analyze the main bottlenecks behind large-scale adoption, and examine emerging solutions poised to redefine secure communication in the quantum era. To ensure a thorough and balanced perspective, this review adopts a systematic methodology for selecting and analyzing the relevant literature. Key terms such as "quantum satellite communication", "QKD protocols", and "entanglement distribution" are used to identify pioneering works and emerging trends. 

As shown in Fig.~\ref{fig_structur}, the article is organized to move quickly from essential background to the main challenge- and outlook-oriented discussion. Section~\ref{sec_principles} provides a concise system-level background, including historical milestones, key metrics, motivating drivers, and enabling technologies. Section~\ref{Sec_theory} briefly summarizes the quantum concepts needed to interpret later sections. Section~\ref{sec_taxonomay} provides a taxonomy of QSC. Section~\ref{Sec_arch} presents current architectures, such as LEO constellations and hybrid quantum-classical networks. Technical and operational challenges, from atmospheric losses to payload limitations, are discussed in Section~\ref{challenge}. Section~\ref{Sec_stateOfArt} highlights recent breakthroughs in protocols, turbulence mitigation, and AI-driven optimizations. Exploring future directions in the quantum Internet and space-based quantum computing is included in Section~\ref{Sec_future}. Finally, we summarize key insights and underscore the transformative potential of QSC in Section~\ref{Sec_conc}.
\begin{table*}[ht!]
 \centering
 {\section*{Acronyms}}
 \resizebox{\textwidth}{!}{%
 \begin{tabular}{llllll}
 \toprule
 \textbf{Acronym} & \textbf{Definition} & \textbf{Acronym} & \textbf{Definition} & \textbf{Acronym} & \textbf{Definition} \\
 \midrule
 \textbf{AO} & adaptive optics & \textbf{ARC} & automated repeater chains & \textbf{AWGN} & additive white Gaussian noise \\
\textbf{CNNs} & convolutional neural networks & \textbf{CV} & continuous variables & \textbf{CV-QKD} & continuous-variable QKD \\
\textbf{DV} & discrete variables & \textbf{EM} & expectation maximization & \textbf{EPR} & Einstein-Podolsky-Rosen \\
\textbf{FDI-QKD} & Faraday invariant QKD & \textbf{FSO} & free space optical & \textbf{GEO} & geostationary earth orbit \\
\textbf{GMM} & Gaussian mixture model & \textbf{GPS} & global positioning systems & \textbf{LEO} & low earth orbit \\
\textbf{LoS} & line-of-sight & \textbf{MDI} & measurement device independent & \textbf{MDI-QKD} & measurement device-independent QKD \\
\textbf{MDI-QSDC} & measurement-device-independent QSDC & \textbf{MEC} & mobile edge computing & \textbf{MEO} & medium earth orbit \\
\textbf{MILP} & mixed-integer linear program & \textbf{ML} & machine learning & \textbf{OAM} & orbital angular momentum \\
\textbf{OPI-QSDC} & one-photon-interference QSDC & \textbf{PDT} & probability distribution of transmittance & \textbf{PLS} & physical layer security \\
\textbf{PM-MDI QKD} & phase-matching MDI-QKD & \textbf{QBER} & quantum bit error rate & \textbf{QCS} & quantum clock synchronization \\
\textbf{QEC} & quantum error correction & \textbf{QKD} & quantum key distribution & \textbf{QMs} & quantum memories \\
\textbf{QPUs} & quantum processing units & \textbf{QSC} & quantum satellite communications & \textbf{QSDC} & quantum secure direct communication \\
\textbf{QuEsat} & ground-satellite network & \textbf{SDN} & software-defined networking & \textbf{SED} & satellite-assisted entanglement distribution \\
\textbf{Si-SPADs} & space-ready single photon avalanche diodes & \textbf{SKR} & secret quantum key rate & \textbf{SNR} & signal to noise ratio \\
\textbf{SPDC} & spontaneous parametric down-conversion & \textbf{SWaP} & size weight and power & \textbf{SED} & satellite-assisted entanglement distribution \\
\textbf{Si-SPADs} & space-ready single photon avalanche diodes & \textbf{SKR} & secret quantum key rate & \textbf{SNR} & signal to noise ratio \\
\textbf{SPDC} & spontaneous parametric down-conversion & \textbf{SWaP} & size weight and power & & \\
 \\
 \bottomrule
 \end{tabular}
 }
\end{table*}
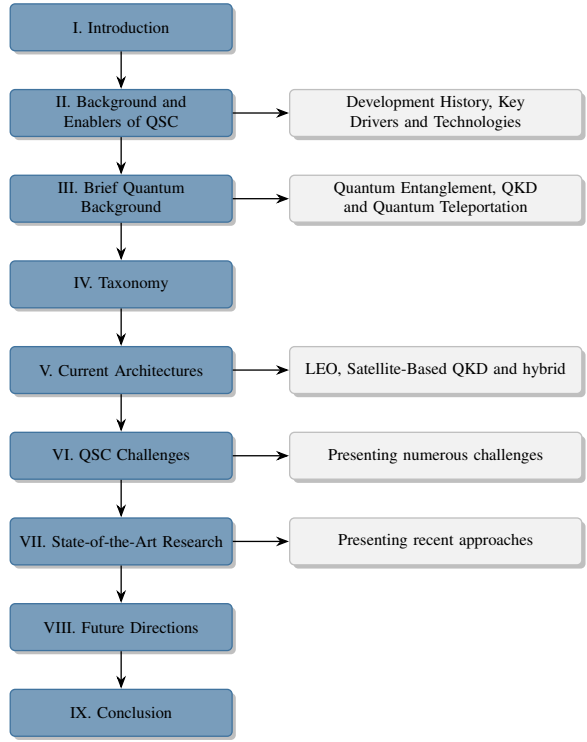
\begin{figure}[ht!]
    \definecolor{leo-blue}{RGB}{64,115,158}

\tikzset{
  block/.style = {rectangle, rounded corners=3pt, draw=leo-blue, fill=leo-blue!70, thick, minimum height=1cm, text width=4.5cm, align=center, drop shadow},
  subblock/.style = {rectangle, rounded corners=3pt, draw=gray!50, fill=gray!10, thick, minimum height=1cm, text width=6cm, align=center, drop shadow},
  arrow/.style = {-{Stealth[scale=1.2]}, thick},
  node distance=0.7cm
}

\resizebox{0.62\width}{!}{
\begin{tikzpicture}[node distance=0.8cm and 1.2cm]

\node[block] (intro) {I. Introduction};

\node[block, below=of intro] (principle) {II. Background and Enablers of QSC};
\node[subblock, right=of principle] (p1) {Development History, Key Drivers and Technologies};

\node[block, below=of principle] (theory) {III. Brief Quantum Background};
\node[subblock, right=of theory] (t1) {Quantum Entanglement, QKD and Quantum Teleportation};

\node[block, below=of theory] (taxonomy) {IV. Taxonomy};

\node[block, below=of taxonomy] (architecture) {V. Current Architectures};
\node[subblock, right=of architecture] (a1) {LEO, Satellite-Based QKD and hybrid};

\node[block, below=of architecture] (challenges) {VI. QSC Challenges};
\node[subblock, right=of challenges] (c1) {Presenting numerous challenges};

\node[block, below=of challenges] (state) {VII. State-of-the-Art Research};
\node[subblock, right=of state] (s1) {Presenting recent approaches};

\node[block, below=of state] (future) {VIII. Future Directions};

\node[block, below=of future] (conclusion) {IX. Conclusion};

% Draw arrows
\foreach \i/\j in {intro/principle, principle/theory, theory/taxonomy, taxonomy/architecture, architecture/challenges,challenges/state,state/future,future/conclusion}
  \draw[arrow] (\i) -- (\j);

% \foreach \main/\sub in {principle/p1, theory/t1, architecture/a1, literature/l1, challenges/c2, state/s1}
%   \draw[arrow] (\main) -- (\sub);

\draw[arrow] (principle) -- (p1);
\draw[arrow] (theory) -- (t1);
\draw[arrow] (architecture) -- (a1);
\draw[arrow] (challenges) -- (c1);
\draw[arrow] (state) -- (s1);

\end{tikzpicture}
}
    \caption{Paper structure}
    \label{fig_structur}
\end{figure}

\section{Background and Enablers of QSC}
\label{sec_principles}
This section provides only the essential background needed for the rest of the review. Rather than expanding introductory material in depth, it briefly summarizes the historical milestones, system metrics, motivating drivers, and enabling technologies that frame the later discussion on challenges, recent progress, and future directions. 

\subsection{Development History of QSC}
The development of QSC has been shaped by several significant milestones. The launch of the Micius satellite in 2016 by China was a pivotal moment, which demonstrates the feasibility of quantum teleportation and QKD over 1,200 km \cite{koushik2020literature,chapman2022paving}. By successfully using entanglement distribution and quantum state teleportation, this satellite laid crucial groundwork for the next generation of QSC technologies. Since then, further missions have continued to build upon these early successes. The underlying architecture of QSC is typically structured into three core segments \cite{sidhu2021advances, Orsucci2024AssessmentOP}:
\begin{enumerate}
    \item Space segment: This includes satellites equipped with quantum transponders and antennas for transmitting quantum states. Low Earth Orbit (LEO) satellites are preferred because of their balance between coverage and signal strength.
    \item Ground segment: This segment handles the reception and processing of quantum signals, comprising ground stations and control centers. The main function is to ensure the integrity of the quantum states and manage the distribution of cryptographic keys.
    \item Network segment: This layer integrates space and ground components, enabling the creation of a global quantum network. It facilitates the exchange of quantum keys and ensures secure communication.
\end{enumerate}

\subsection{Key Drivers Motivating QSC}
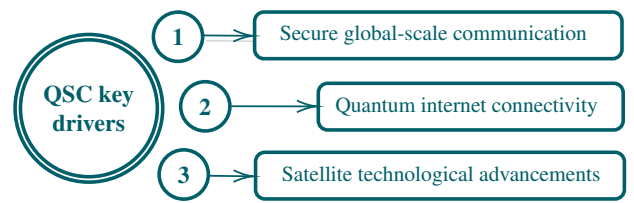
\begin{figure}[ht!]
    \definecolor{colorgreen}{HTML}{005F69}

\tikzset{every picture/.style={line width=0.75pt}} %set default line width to 0.75pt        
\centering
\resizebox{0.9\width}{!}{
\begin{tikzpicture}[x=0.75pt,y=0.75pt,yscale=-1,xscale=1]

%Straight Lines [id:da6029492045895763] 
\draw [color=colorgreen  ,draw opacity=1 , drop shadow][line width=0.75]    (185.5,101.75) -- (212.6,101.98) ;
\draw [shift={(214.6,102)}, rotate = 180.49] [color=colorgreen  ,draw opacity=1 ][line width=0.75]    (10.93,-3.29) .. controls (6.95,-1.4) and (3.31,-0.3) .. (0,0) .. controls (3.31,0.3) and (6.95,1.4) .. (10.93,3.29)   ;
%Rounded Rect [id:dp634489086037205] 
\draw  [color=colorgreen  ,draw opacity=1 ][line width=1.5]  (215.47,93.93) .. controls (215.47,91.06) and (217.79,88.73) .. (220.67,88.73) -- (415.13,88.73) .. controls (418.01,88.73) and (420.33,91.06) .. (420.33,93.93) -- (420.33,109.53) .. controls (420.33,112.41) and (418.01,114.73) .. (415.13,114.73) -- (220.67,114.73) .. controls (217.79,114.73) and (215.47,112.41) .. (215.47,109.53) -- cycle ;
%Rounded Rect [id:dp57350915812288] 
\draw  [color=colorgreen  ,draw opacity=1 ][line width=1.5]  (251,133.67) .. controls (251,130.79) and (253.33,128.47) .. (256.2,128.47) -- (415.8,128.47) .. controls (418.67,128.47) and (421,130.79) .. (421,133.67) -- (421,149.27) .. controls (421,152.14) and (418.67,154.47) .. (415.8,154.47) -- (256.2,154.47) .. controls (253.33,154.47) and (251,152.14) .. (251,149.27) -- cycle ;
%Rounded Rect [id:dp38928737957111204] 
\draw  [color=colorgreen  ,draw opacity=1 ][line width=1.5]  (216,172.13) .. controls (216,169.26) and (218.33,166.93) .. (221.2,166.93) -- (416.47,166.93) .. controls (419.34,166.93) and (421.67,169.26) .. (421.67,172.13) -- (421.67,187.73) .. controls (421.67,190.61) and (419.34,192.93) .. (416.47,192.93) -- (221.2,192.93) .. controls (218.33,192.93) and (216,190.61) .. (216,187.73) -- cycle ;
%Straight Lines [id:da6750597408368282] 
\draw [color=colorgreen  ,draw opacity=1 ][line width=0.75]    (202.5,141.25) -- (248,141.33) ;
\draw [shift={(250,141.33)}, rotate = 180.1] [color=colorgreen  ,draw opacity=1 ][line width=0.75]    (10.93,-3.29) .. controls (6.95,-1.4) and (3.31,-0.3) .. (0,0) .. controls (3.31,0.3) and (6.95,1.4) .. (10.93,3.29)   ;
%Straight Lines [id:da02908531313546625] 
\draw [color=colorgreen  ,draw opacity=1 ][line width=0.75]    (189.4,180) -- (212.67,180.31) ;
\draw [shift={(214.67,180.33)}, rotate = 180.76] [color=colorgreen  ,draw opacity=1 ][line width=0.75]    (10.93,-3.29) .. controls (6.95,-1.4) and (3.31,-0.3) .. (0,0) .. controls (3.31,0.3) and (6.95,1.4) .. (10.93,3.29)   ;

% Text Node
\draw  [color=colorgreen  ,draw opacity=1 ][line width=1.5]   (123.08, 142.4) circle [x radius= 38.16, y radius= 38.16]   (123.08, 142.4) circle [x radius= 41.16, y radius= 41.16]  ;
\draw (123.08,142.4) node  [font=\normalsize,color=colorgreen  ,opacity=1 ] [align=left] {\begin{minipage}[lt]{43.03pt}\setlength\topsep{0pt}
\begin{center}
\textbf{QSC key }\\\textbf{drivers}
\end{center}

\end{minipage}};
% Text Node
\draw  [color=colorgreen  ,draw opacity=1 ][line width=1.5]   (172.77, 102.5) circle [x radius= 13.73, y radius= 13.73]   ;
\draw (172.77,102.5) node  [font=\normalsize,color=colorgreen] [align=left] {\begin{minipage}[lt]{8.67pt}\setlength\topsep{0pt}
\begin{center}
\textbf{{1}}
\end{center}

\end{minipage}};
% Text Node
\draw (228,94.93) node [anchor=north west][inner sep=0.75pt]  [font=\small,color=colorgreen  ,opacity=1 ] [align=center] {Secure global-scale communication};
% Text Node
\draw (259,134.93) node [anchor=north west][inner sep=0.75pt]  [font=\small,color=colorgreen  ,opacity=1 ] [align=left] {Quantum internet connectivity};
% Text Node
\draw (319.57,180.27) node  [font=\small,color=colorgreen  ,opacity=1 ] [align=left] {Satellite technological advancements};
% Text Node
\draw  [color=colorgreen  ,draw opacity=1 ][line width=1.5]   (187.97, 141.3) circle [x radius= 13.73, y radius= 13.73]   ;
\draw (187.97,141.3) node  [font=\normalsize,color=colorgreen  ,opacity=1 ] [align=left] {\begin{minipage}[lt]{8.67pt}\setlength\topsep{0pt}
\begin{center}
\textbf{2}
\end{center}

\end{minipage}};
% Text Node
\draw  [color=colorgreen  ,draw opacity=1 ][line width=1.5]   (175.97, 179.3) circle [x radius= 13.73, y radius= 13.73]   ;
\draw (175.97,179.3) node  [font=\normalsize,color=colorgreen  ,opacity=1 ] [align=left] {\begin{minipage}[lt]{8.67pt}\setlength\topsep{0pt}
\begin{center}
\textbf{3}
\end{center}

\end{minipage}};

\end{tikzpicture}
}
    \caption{QSC key drivers}
    \label{fig_key}
\end{figure}

QSC is driven by several key factors as illustrated in Fig.~\ref{fig_key} \cite{sidhu2021advances,lai2023application}. Foremost among these is the growing demand for secure communication on a global scale. Conventional communication methods face challenges in ensuring absolute security, while QSC takes advantage of the principles of quantum mechanics, such as the no-cloning theorem and quantum entanglement, to enable encryption that is, in theory, unbreakable. The second major driver is the vision of establishing a global quantum Internet. Just as classical satellites form the backbone of global communication networks, quantum satellites can serve as the core nodes within a future quantum network, supporting the distribution of quantum keys and entangled particles over long-distance areas \cite{huang2020quantum}. Lastly, advances in satellite technology and quantum science have made QSC more feasible, where the development of smaller and more efficient satellites, combined with breakthroughs in quantum optics and error correction techniques, has moved this vision of QSC closer to practical realization.

\subsection{Key Technologies in QSC}
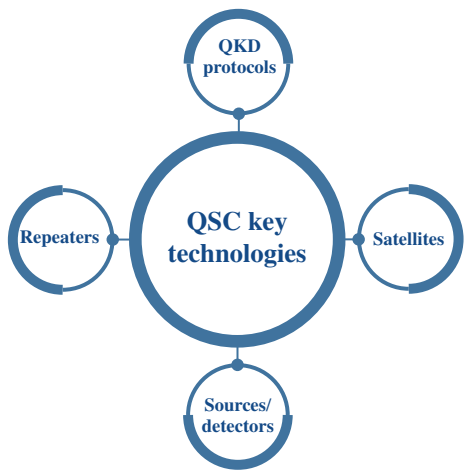
\begin{figure}[ht!]
    \definecolor{leo-blue}{RGB}{64,115,158}

\tikzset{every picture/.style={line width=0.75pt}} %set default line width to 0.75pt        
\centering
\begin{tikzpicture}[x=0.75pt,y=0.75pt,yscale=-1,xscale=1]

%Shape: Arc [id:dp5279760914353724] 
\draw  [draw opacity=0][line width=1.5]  (218.1,65.8) .. controls (218.1,65.8) and (218.1,65.8) .. (218.1,65.8) .. controls (218.1,79.61) and (206.91,90.8) .. (193.1,90.8) .. controls (179.29,90.8) and (168.1,79.61) .. (168.1,65.8) -- (193.1,65.8) -- cycle ; \draw  [color=leo-blue  ,draw opacity=1 ][line width=1.5]  (218.1,65.8) .. controls (218.1,65.8) and (218.1,65.8) .. (218.1,65.8) .. controls (218.1,79.61) and (206.91,90.8) .. (193.1,90.8) .. controls (179.29,90.8) and (168.1,79.61) .. (168.1,65.8) ;  
%Shape: Arc [id:dp25404573115839413] 
\draw  [draw opacity=0][line width=3.75]  (168.1,65.8) .. controls (168.1,65.8) and (168.1,65.8) .. (168.1,65.8) .. controls (168.1,51.99) and (179.29,40.8) .. (193.1,40.8) .. controls (206.91,40.8) and (218.1,51.99) .. (218.1,65.8) -- (193.1,65.8) -- cycle ; \draw  [color=leo-blue  ,draw opacity=1 ][line width=3.75]  (168.1,65.8) .. controls (168.1,65.8) and (168.1,65.8) .. (168.1,65.8) .. controls (168.1,51.99) and (179.29,40.8) .. (193.1,40.8) .. controls (206.91,40.8) and (218.1,51.99) .. (218.1,65.8) ;  
%Straight Lines [id:da14574276225110006] 
\draw [color=leo-blue  ,draw opacity=1 ]   (193.3,90.95) -- (193.3,103.75) ;
%Flowchart: Connector [id:dp39092324464131956] 
\draw  [color=leo-blue  ,draw opacity=1 ][fill=leo-blue  ,fill opacity=1 ] (190.68,90.95) .. controls (190.68,89.5) and (191.85,88.32) .. (193.3,88.32) .. controls (194.75,88.32) and (195.93,89.5) .. (195.93,90.95) .. controls (195.93,92.4) and (194.75,93.57) .. (193.3,93.57) .. controls (191.85,93.57) and (190.68,92.4) .. (190.68,90.95) -- cycle ;
%Shape: Arc [id:dp370294010033257] 
\draw  [draw opacity=0][line width=1.5]  (278.58,178.78) .. controls (278.58,178.78) and (278.58,178.78) .. (278.58,178.78) .. controls (264.77,178.78) and (253.58,167.58) .. (253.58,153.78) .. controls (253.58,139.97) and (264.77,128.78) .. (278.57,128.78) -- (278.58,153.78) -- cycle ; \draw  [color=leo-blue  ,draw opacity=1 ][line width=1.5]  (278.58,178.78) .. controls (278.58,178.78) and (278.58,178.78) .. (278.58,178.78) .. controls (264.77,178.78) and (253.58,167.58) .. (253.58,153.78) .. controls (253.58,139.97) and (264.77,128.78) .. (278.57,128.78) ;  
%Shape: Arc [id:dp8853757757811207] 
\draw  [draw opacity=0][line width=3.75]  (278.58,128.78) .. controls (278.58,128.78) and (278.58,128.78) .. (278.58,128.78) .. controls (292.38,128.78) and (303.58,139.97) .. (303.58,153.78) .. controls (303.58,167.58) and (292.38,178.78) .. (278.58,178.78) -- (278.58,153.78) -- cycle ; \draw  [color=leo-blue  ,draw opacity=1 ][line width=3.75]  (278.58,128.78) .. controls (278.58,128.78) and (278.58,128.78) .. (278.58,128.78) .. controls (292.38,128.78) and (303.58,139.97) .. (303.58,153.78) .. controls (303.58,167.58) and (292.38,178.78) .. (278.58,178.78) ;  
%Straight Lines [id:da7178028876330285] 
\draw [color=leo-blue  ,draw opacity=1 ]   (253.43,153.98) -- (240.63,153.98) ;
%Flowchart: Connector [id:dp7895518090920821] 
\draw  [color=leo-blue  ,draw opacity=1 ][fill=leo-blue  ,fill opacity=1 ] (253.43,151.35) .. controls (254.87,151.35) and (256.05,152.53) .. (256.05,153.98) .. controls (256.05,155.42) and (254.87,156.6) .. (253.43,156.6) .. controls (251.98,156.6) and (250.8,155.42) .. (250.8,153.98) .. controls (250.8,152.53) and (251.98,151.35) .. (253.43,151.35) -- cycle ;
%Shape: Arc [id:dp4461357200900151] 
\draw  [draw opacity=0][line width=1.5]  (104.88,129.27) .. controls (104.88,129.27) and (104.88,129.27) .. (104.88,129.27) .. controls (118.68,129.27) and (129.88,140.47) .. (129.88,154.28) .. controls (129.88,168.08) and (118.68,179.28) .. (104.88,179.28) -- (104.88,154.28) -- cycle ; \draw  [color=leo-blue  ,draw opacity=1 ][line width=1.5]  (104.88,129.27) .. controls (104.88,129.27) and (104.88,129.27) .. (104.88,129.27) .. controls (118.68,129.27) and (129.88,140.47) .. (129.88,154.28) .. controls (129.88,168.08) and (118.68,179.28) .. (104.88,179.28) ;  
%Shape: Arc [id:dp9131774698437416] 
\draw  [draw opacity=0][line width=3.75]  (104.88,179.28) .. controls (104.88,179.28) and (104.88,179.28) .. (104.88,179.28) .. controls (91.07,179.28) and (79.88,168.08) .. (79.88,154.28) .. controls (79.88,140.47) and (91.07,129.28) .. (104.87,129.28) -- (104.88,154.28) -- cycle ; \draw  [color=leo-blue  ,draw opacity=1 ][line width=3.75]  (104.88,179.28) .. controls (104.88,179.28) and (104.88,179.28) .. (104.88,179.28) .. controls (91.07,179.28) and (79.88,168.08) .. (79.88,154.28) .. controls (79.88,140.47) and (91.07,129.28) .. (104.87,129.28) ;  
%Straight Lines [id:da8625722262691631] 
\draw [color=leo-blue  ,draw opacity=1 ]   (130.03,154.08) -- (142.82,154.08) ;
%Flowchart: Connector [id:dp45242038684545394] 
\draw  [color=leo-blue  ,draw opacity=1 ][fill=leo-blue  ,fill opacity=1 ] (130.03,156.7) .. controls (128.58,156.7) and (127.4,155.52) .. (127.4,154.08) .. controls (127.4,152.63) and (128.58,151.45) .. (130.03,151.45) .. controls (131.47,151.45) and (132.65,152.63) .. (132.65,154.08) .. controls (132.65,155.52) and (131.47,156.7) .. (130.03,156.7) -- cycle ;
%Shape: Arc [id:dp3153924293036876] 
\draw  [draw opacity=0][line width=1.5]  (167.83,242.12) .. controls (167.83,242.12) and (167.83,242.12) .. (167.83,242.12) .. controls (167.83,228.31) and (179.03,217.12) .. (192.83,217.12) .. controls (206.64,217.12) and (217.83,228.31) .. (217.83,242.12) -- (192.83,242.12) -- cycle ; \draw  [color=leo-blue  ,draw opacity=1 ][line width=1.5]  (167.83,242.12) .. controls (167.83,242.12) and (167.83,242.12) .. (167.83,242.12) .. controls (167.83,228.31) and (179.03,217.12) .. (192.83,217.12) .. controls (206.64,217.12) and (217.83,228.31) .. (217.83,242.12) ;  
%Shape: Arc [id:dp40674497159133693] 
\draw  [draw opacity=0][line width=3.75]  (217.83,242.12) .. controls (217.83,242.12) and (217.83,242.12) .. (217.83,242.12) .. controls (217.83,255.92) and (206.64,267.12) .. (192.83,267.12) .. controls (179.03,267.12) and (167.83,255.92) .. (167.83,242.12) -- (192.83,242.12) -- cycle ; \draw  [color=leo-blue  ,draw opacity=1 ][line width=3.75]  (217.83,242.12) .. controls (217.83,242.12) and (217.83,242.12) .. (217.83,242.12) .. controls (217.83,255.92) and (206.64,267.12) .. (192.83,267.12) .. controls (179.03,267.12) and (167.83,255.92) .. (167.83,242.12) ;  
%Straight Lines [id:da23216569607544457] 
\draw [color=leo-blue  ,draw opacity=1 ]   (192.63,216.97) -- (192.63,204.17) ;
%Flowchart: Connector [id:dp701243743671244] 
\draw  [color=leo-blue  ,draw opacity=1 ][fill=leo-blue  ,fill opacity=1 ] (195.26,216.97) .. controls (195.26,218.42) and (194.08,219.59) .. (192.63,219.59) .. controls (191.18,219.59) and (190.01,218.42) .. (190.01,216.97) .. controls (190.01,215.52) and (191.18,214.34) .. (192.63,214.34) .. controls (194.08,214.34) and (195.26,215.52) .. (195.26,216.97) -- cycle ;

% Text Node
\draw  [color=leo-blue  ,draw opacity=1 ][line width=2.5]   (192.5, 154) circle [x radius= 49.64, y radius= 49.64]   (192.5, 154) circle [x radius= 52.64, y radius= 52.64]  ;
\draw (192.5,154) node [align=center] {\begin{minipage}[lt]{60pt}\setlength\topsep{0pt}
\begin{center}
\textcolor[rgb]{0.08,0.29,0.53}{\textbf{QSC key}}\\\textcolor[rgb]{0.08,0.29,0.53}{\textbf{technologies}}
\end{center}

\end{minipage}};
% Text Node
\draw (193.79,62.89) node  [font=\scriptsize] [align=center] {\begin{minipage}[lt]{32.49pt}\setlength\topsep{0pt}
\begin{center}
\textcolor[rgb]{0.08,0.29,0.53}{\textbf{QKD}}\\\textcolor[rgb]{0.08,0.29,0.53}{\textbf{protocols}}
\end{center}

\end{minipage}};
% Text Node
\draw (278.58,153.78) node  [font=\scriptsize] [align=center] {\begin{minipage}[lt]{31.69pt}\setlength\topsep{0pt}
\begin{center}
\textcolor[rgb]{0.08,0.29,0.53}{\textbf{Satellites}}
\end{center}

\end{minipage}};
% Text Node
\draw (103.38,153.53) node  [font=\scriptsize] [align=center] {\begin{minipage}[lt]{35.67pt}\setlength\topsep{0pt}
\begin{center}
\textcolor[rgb]{0.08,0.29,0.53}{\textbf{Repeaters}}
\end{center}

\end{minipage}};
% Text Node
\draw (192.83,242.12) node  [font=\scriptsize] [align=center] {\begin{minipage}[lt]{32.1pt}\setlength\topsep{0pt}
\begin{center}
\textcolor[rgb]{0.08,0.29,0.53}{\textbf{Sources}/}\\\textcolor[rgb]{0.08,0.29,0.53}{\textbf{detectors}}
\end{center}

\end{minipage}};

\end{tikzpicture}
    \caption{QSC key technologies}
    \label{fig_key_tec}
\end{figure}

An overview of these enabling technologies is provided in Fig.~\ref{fig_key_tec}. A cornerstone of QSC is QKD, which enables two parties to share a secret key securely using the principles of quantum mechanics. This is typically implemented through well-established protocols, such as BB84, which employs weak coherent pulses, and BBM92, which makes use of entangled photon pairs. In this context, satellites serve as essential platforms for transmitting quantum states over long distances, particularly LEO. This is achieved through QKD protocols such as the BB84 protocol, which uses weak coherent pulses, or the BBM92 protocol, which relies on entangled photon pairs. Satellites play a crucial role in this process, which use LEO satellites for quantum state transmission. To further extend the range of QKD, quantum repeaters are employed. These devices help maintain entanglement across vast distances, thereby maintaining the scalability of quantum networks. Another vital component of QSC lies in the development of compact and robust quantum sources and detectors that can operate in a harsh space environment \cite{sodagari2023integrating}.

\subsection{Quantum Repeaters and Entanglement Swapping}
\label{sec:quantum_repeaters}

Quantum repeaters are the cornerstone of any long-distance quantum communication system, including quantum satellite networks. They overcome the exponential photon loss and decoherence challenges that limit the range of direct quantum links. A quantum repeater works by dividing the communication channel into shorter segments, distributing entanglement across these segments, and then performing a process known as entanglement swapping to extend the entanglement over the entire distance. This architecture enables scalable, high-fidelity quantum links that are vital for building a global quantum internet.

Entanglement swapping allows two independent entangled pairs to be converted into a new pair of entangled nodes, even if these nodes have never directly interacted. In practice, this is performed by a joint Bell-state measurement at an intermediate node, which effectively "swaps" the entanglement and extends the quantum correlation over a longer distance. This mechanism is essential for enabling secure quantum key distribution (QKD), quantum teleportation, and distributed quantum computing over continental or global scales.

Recent advancements in quantum repeater technologies have significantly improved their practicality and efficiency. For instance, all-photonic quantum repeaters eliminate the need for quantum memories by using large-scale photonic cluster states, making the system more robust and easier to implement. Similarly, Zeno-based quantum repeaters exploit the quantum Zeno effect to reduce circuit complexity and enhance the stability of entanglement distribution over noisy channels. These innovations, alongside advances in ultrabright photon sources and integrated photonic platforms, are paving the way for next-generation satellite-assisted quantum networks.

In the context of quantum satellite communication, integrating advanced repeater nodes with satellite and terrestrial segments promises a seamless global quantum network. These hybrid architectures can deliver reliable and efficient entanglement distribution across thousands of kilometers, positioning quantum repeaters as indispensable components for future quantum internet infrastructures.

\section{Brief Quantum Background}
\label{Sec_theory}
\subsection{Quantum Entanglement and Its Role in QSC}
\begin{figure*}[htbp]
    \centering
    \includegraphics[width=1.0\linewidth]{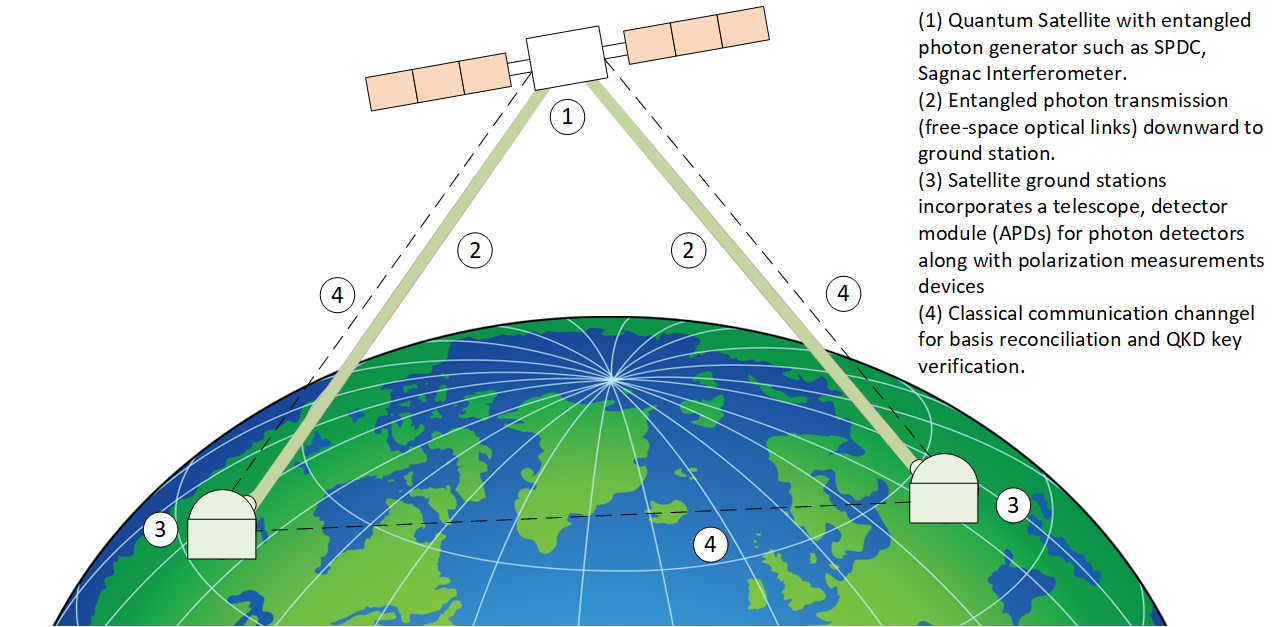}
    \caption{Satellite quantum entanglement illustration}
    \label{Entangled}
\end{figure*}

Quantum entanglement is the key resource that allows two distant nodes to share correlated quantum states in a way that supports secure communication and distributed quantum processing. For QSC, its importance is primarily practical: entanglement enables long-distance key establishment, supports teleportation-based operations, and offers security advantages because interception attempts disturb the shared state \cite{atlas2024observation,singh2024quick}. In satellite settings, entangled photons may be generated onboard or distributed across space-to-ground links, making entanglement distribution a central systems challenge rather than a purely theoretical topic. Recent experimental progress confirms that such links are feasible over very long distances, motivating the architectural and implementation issues discussed later in this review \cite{gonzalez2024satellite}.

\subsection{Quantum Key Distribution (QKD)}
QKD is the most mature application area in QSC and the main reason satellite quantum links attract immediate practical interest. Its value in this review lies less in protocol derivations than in how protocol choices shape system requirements, security assumptions, and deployment constraints. Representative families include prepare-and-measure schemes such as BB84, entanglement-based approaches such as E91 and BBM92, and more implementation-aware variants such as decoy-state, measurement-device-independent, and continuous-variable QKD. In satellite scenarios, these protocols may be deployed over uplink, downlink, or inter-satellite channels, each imposing different trade-offs in loss, pointing accuracy, detector performance, and trust assumptions. For that reason, QKD is best understood here as the bridge between quantum principles and the practical bottlenecks analyzed in the challenge sections.

\subsection{Quantum Teleportation}
Quantum teleportation enables the transfer of an unknown quantum state by combining pre-shared entanglement with classical communication. In the context of this review, its importance is that it illustrates how satellite platforms may eventually support advanced quantum networking functions beyond direct key exchange. Teleportation experiments over space links have already demonstrated the feasibility of distributing quantum information across long distances, but they also expose strict requirements on link stability, entanglement fidelity, timing, and onboard resources.

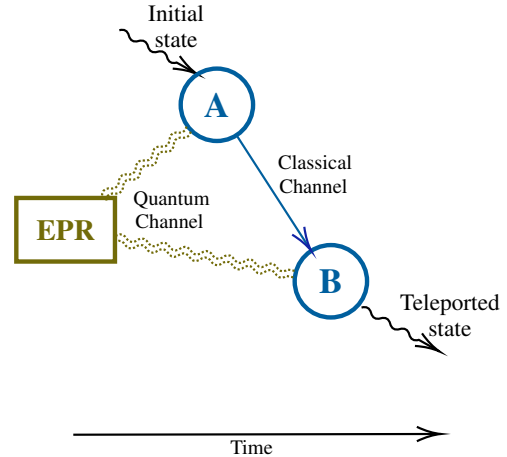
\begin{figure}[htbp]
    \centering
    \definecolor{colorblue}{HTML}{005F9E}

\tikzset{every picture/.style={line width=0.75pt}} %set default line width to 0.75pt
\centering
\resizebox{1.1\width}{!}{
\begin{tikzpicture}[x=0.75pt,y=0.75pt,yscale=-1,xscale=1]

%Straight Lines [id:da1178559906865988] 
\draw    (194.5,152) .. controls (196.77,151.37) and (198.22,152.18) .. (198.86,154.45) .. controls (199.49,156.72) and (200.94,157.54) .. (203.21,156.91) .. controls (205.48,156.28) and (206.93,157.09) .. (207.57,159.36) .. controls (208.2,161.63) and (209.65,162.45) .. (211.92,161.82) .. controls (214.19,161.19) and (215.64,162) .. (216.28,164.27) .. controls (216.91,166.54) and (218.37,167.36) .. (220.64,166.73) -- (221.29,167.09) -- (228.26,171.02) ;
\draw [shift={(230,172)}, rotate = 209.4] [color={rgb, 255:red, 0; green, 0; blue, 0 }  ][line width=0.75]    (10.93,-3.29) .. controls (6.95,-1.4) and (3.31,-0.3) .. (0,0) .. controls (3.31,0.3) and (6.95,1.4) .. (10.93,3.29)   ;
%Straight Lines [id:da41174509466996434] 
\draw    (84,25.5) .. controls (86.27,24.87) and (87.72,25.68) .. (88.36,27.95) .. controls (88.99,30.22) and (90.44,31.04) .. (92.71,30.41) .. controls (94.98,29.78) and (96.43,30.59) .. (97.07,32.86) .. controls (97.7,35.13) and (99.15,35.95) .. (101.42,35.32) .. controls (103.69,34.69) and (105.14,35.5) .. (105.78,37.77) .. controls (106.41,40.04) and (107.87,40.86) .. (110.14,40.23) -- (110.79,40.59) -- (117.76,44.52) ;
\draw [shift={(119.5,45.5)}, rotate = 209.4] [color={rgb, 255:red, 0; green, 0; blue, 0 }  ][line width=0.75]    (10.93,-3.29) .. controls (6.95,-1.4) and (3.31,-0.3) .. (0,0) .. controls (3.31,0.3) and (6.95,1.4) .. (10.93,3.29)   ;
%Straight Lines [id:da011455667511670597] 
\draw    (63,210) -- (228.67,209.67) ;
\draw [shift={(230.67,209.67)}, rotate = 179.89] [color={rgb, 255:red, 0; green, 0; blue, 0 }  ][line width=0.75]    (10.93,-3.29) .. controls (6.95,-1.4) and (3.31,-0.3) .. (0,0) .. controls (3.31,0.3) and (6.95,1.4) .. (10.93,3.29)   ;

% Text Node
\draw  [color=colorblue  ,draw opacity=1 ][line width=1.5]   (128.51, 59.5) circle [x radius= 16.51, y radius= 16.51]   ;
\draw (128.51,59.5) node  [font=\large,color=colorblue  ,opacity=1 ] [align=left] {\textbf{{A}}};
% Text Node
\draw  [color=colorblue  ,draw opacity=1 ][line width=1.5]   (180.51, 140) circle [x radius= 16.51, y radius= 16.51]   ;
\draw (180.51,140) node  [font=\large,color=colorblue  ,opacity=1 ] [align=left] {\textbf{{B}}};
% Text Node
\draw  [color={rgb, 255:red, 114; green, 107; blue, 9 }  ,draw opacity=1 ][line width=1.5]   (37.01,102) -- (82.01,102) -- (82.01,131) -- (37.01,131) -- cycle  ;
\draw (59.51,116.5) node  [font=\normalsize,color={rgb, 255:red, 114; green, 107; blue, 9 }  ,opacity=1 ] [align=left] {\textbf{{EPR}}};
% Text Node
\draw (85.83,12.33) node [anchor=north west][inner sep=0.75pt]  [font=\footnotesize] [align=left] {\begin{minipage}[lt]{33.93pt}\setlength\topsep{0pt}
\begin{center}
{\footnotesize Initial state}
\end{center}

\end{minipage}};
% Text Node
\draw (210,144.17) node [anchor=north west][inner sep=0.75pt]  [font=\footnotesize] [align=left] {\begin{minipage}[lt]{35.02pt}\setlength\topsep{0pt}
\begin{center}
{\footnotesize Teleported }\\{\footnotesize state}
\end{center}

\end{minipage}};
% Text Node
\draw (130.67,210.83) node [anchor=north west][inner sep=0.75pt]  [font=\scriptsize] [align=left] {\begin{minipage}[lt]{18.32pt}\setlength\topsep{0pt}
\begin{center}
Time
\end{center}

\end{minipage}};
% Text Node
\draw (155,76.83) node [anchor=north west][inner sep=0.75pt]  [font=\scriptsize] [align=left] {\begin{minipage}[lt]{24.1pt}\setlength\topsep{0pt}
\begin{center}
{\scriptsize Classical }\\{\scriptsize Channel}
\end{center}

\end{minipage}};
% Text Node
\draw (89,93) node [anchor=north west][inner sep=0.75pt]  [font=\scriptsize] [align=left] {\begin{minipage}[lt]{24.65pt}\setlength\topsep{0pt}
\begin{center}
{\scriptsize Quantum }\\{\scriptsize Channel}
\end{center}

\end{minipage}};
% Connection
\draw [color=colorblue  ,draw opacity=1 ]   (137.47,73.37) -- (170.46,124.45) ;
\draw [shift={(171.55,126.13)}, rotate = 237.14] [color={rgb, 255:red, 18; green, 35; blue, 165 }  ,draw opacity=1 ][line width=0.75]    (10.93,-3.29) .. controls (6.95,-1.4) and (3.31,-0.3) .. (0,0) .. controls (3.31,0.3) and (6.95,1.4) .. (10.93,3.29)   ;
% Connection
\draw [color={rgb, 255:red, 114; green, 107; blue, 9 }  ,draw opacity=1 ] [dash pattern={on 0.75pt off 0.75pt}]  (76.11,100.84) .. controls (76.33,98.49) and (77.61,97.43) .. (79.96,97.66) .. controls (82.31,97.88) and (83.59,96.82) .. (83.81,94.47) .. controls (84.04,92.12) and (85.32,91.06) .. (87.67,91.29) .. controls (90.02,91.52) and (91.3,90.46) .. (91.52,88.11) .. controls (91.75,85.76) and (93.03,84.7) .. (95.38,84.92) .. controls (97.73,85.15) and (99.01,84.09) .. (99.23,81.74) .. controls (99.46,79.39) and (100.74,78.33) .. (103.09,78.55) .. controls (105.44,78.78) and (106.72,77.72) .. (106.94,75.37) .. controls (107.17,73.02) and (108.45,71.96) .. (110.8,72.18) .. controls (113.15,72.41) and (114.43,71.35) .. (114.65,69) -- (114.83,68.86) -- (114.83,68.86)(78.02,103.16) .. controls (78.24,100.81) and (79.52,99.75) .. (81.87,99.97) .. controls (84.22,100.2) and (85.5,99.14) .. (85.73,96.79) .. controls (85.95,94.44) and (87.23,93.38) .. (89.58,93.6) .. controls (91.93,93.83) and (93.21,92.77) .. (93.43,90.42) .. controls (93.66,88.07) and (94.94,87.01) .. (97.29,87.23) .. controls (99.64,87.46) and (100.92,86.4) .. (101.14,84.05) .. controls (101.37,81.7) and (102.65,80.64) .. (105,80.87) .. controls (107.35,81.09) and (108.63,80.03) .. (108.85,77.68) .. controls (109.08,75.33) and (110.36,74.27) .. (112.71,74.5) .. controls (115.06,74.72) and (116.34,73.66) .. (116.56,71.31) -- (116.74,71.17) -- (116.74,71.17) ;
% Connection
\draw [color={rgb, 255:red, 114; green, 107; blue, 9 }  ,draw opacity=1 ] [dash pattern={on 0.75pt off 0.75pt}]  (82.29,119.4) .. controls (84.24,118.08) and (85.88,118.4) .. (87.2,120.35) .. controls (88.52,122.3) and (90.16,122.62) .. (92.11,121.3) .. controls (94.06,119.99) and (95.7,120.31) .. (97.02,122.26) .. controls (98.34,124.21) and (99.98,124.53) .. (101.93,123.21) .. controls (103.88,121.89) and (105.52,122.21) .. (106.84,124.16) .. controls (108.15,126.11) and (109.79,126.43) .. (111.74,125.12) .. controls (113.69,123.8) and (115.33,124.12) .. (116.65,126.07) .. controls (117.97,128.02) and (119.61,128.34) .. (121.56,127.02) .. controls (123.51,125.71) and (125.15,126.03) .. (126.47,127.98) .. controls (127.79,129.93) and (129.43,130.25) .. (131.38,128.93) .. controls (133.33,127.61) and (134.97,127.93) .. (136.28,129.88) .. controls (137.6,131.83) and (139.24,132.15) .. (141.19,130.84) .. controls (143.14,129.52) and (144.78,129.84) .. (146.1,131.79) .. controls (147.42,133.74) and (149.06,134.06) .. (151.01,132.74) .. controls (152.96,131.43) and (154.6,131.75) .. (155.92,133.7) .. controls (157.24,135.65) and (158.88,135.97) .. (160.83,134.65) -- (164.59,135.38) -- (164.59,135.38)(81.72,122.34) .. controls (83.67,121.03) and (85.31,121.35) .. (86.63,123.3) .. controls (87.95,125.25) and (89.59,125.57) .. (91.54,124.25) .. controls (93.49,122.93) and (95.13,123.25) .. (96.45,125.2) .. controls (97.76,127.15) and (99.4,127.47) .. (101.35,126.16) .. controls (103.3,124.84) and (104.94,125.16) .. (106.26,127.11) .. controls (107.58,129.06) and (109.22,129.38) .. (111.17,128.06) .. controls (113.12,126.75) and (114.76,127.07) .. (116.08,129.02) .. controls (117.4,130.97) and (119.04,131.29) .. (120.99,129.97) .. controls (122.94,128.65) and (124.58,128.97) .. (125.9,130.92) .. controls (127.21,132.87) and (128.85,133.19) .. (130.8,131.87) .. controls (132.75,130.56) and (134.39,130.88) .. (135.71,132.83) .. controls (137.03,134.78) and (138.67,135.1) .. (140.62,133.78) .. controls (142.57,132.46) and (144.21,132.78) .. (145.53,134.73) .. controls (146.85,136.68) and (148.49,137) .. (150.44,135.69) .. controls (152.39,134.37) and (154.03,134.69) .. (155.35,136.64) .. controls (156.66,138.59) and (158.3,138.91) .. (160.25,137.59) -- (164.01,138.32) -- (164.01,138.32) ;

\end{tikzpicture}
}
    \caption{Quantum teleportation illustration}
    \label{teleportation}
\end{figure}

Satellite teleportation studies further show how future QSC systems could evolve toward repeater-assisted networking, distributed sensing, and eventually space-supported quantum computing \cite{qiu2024deterministic, ren2017ground, chou2023satellite}. At the same time, they make clear that realizing those visions depends on overcoming many of the same engineering barriers discussed later in the paper, especially channel loss, synchronization complexity, and robust entanglement distribution.

\section{Taxonomy}
\label{sec_taxonomay}
\begin{figure*}[ht!]
\begin{center}
\includegraphics[width=0.9\textwidth]{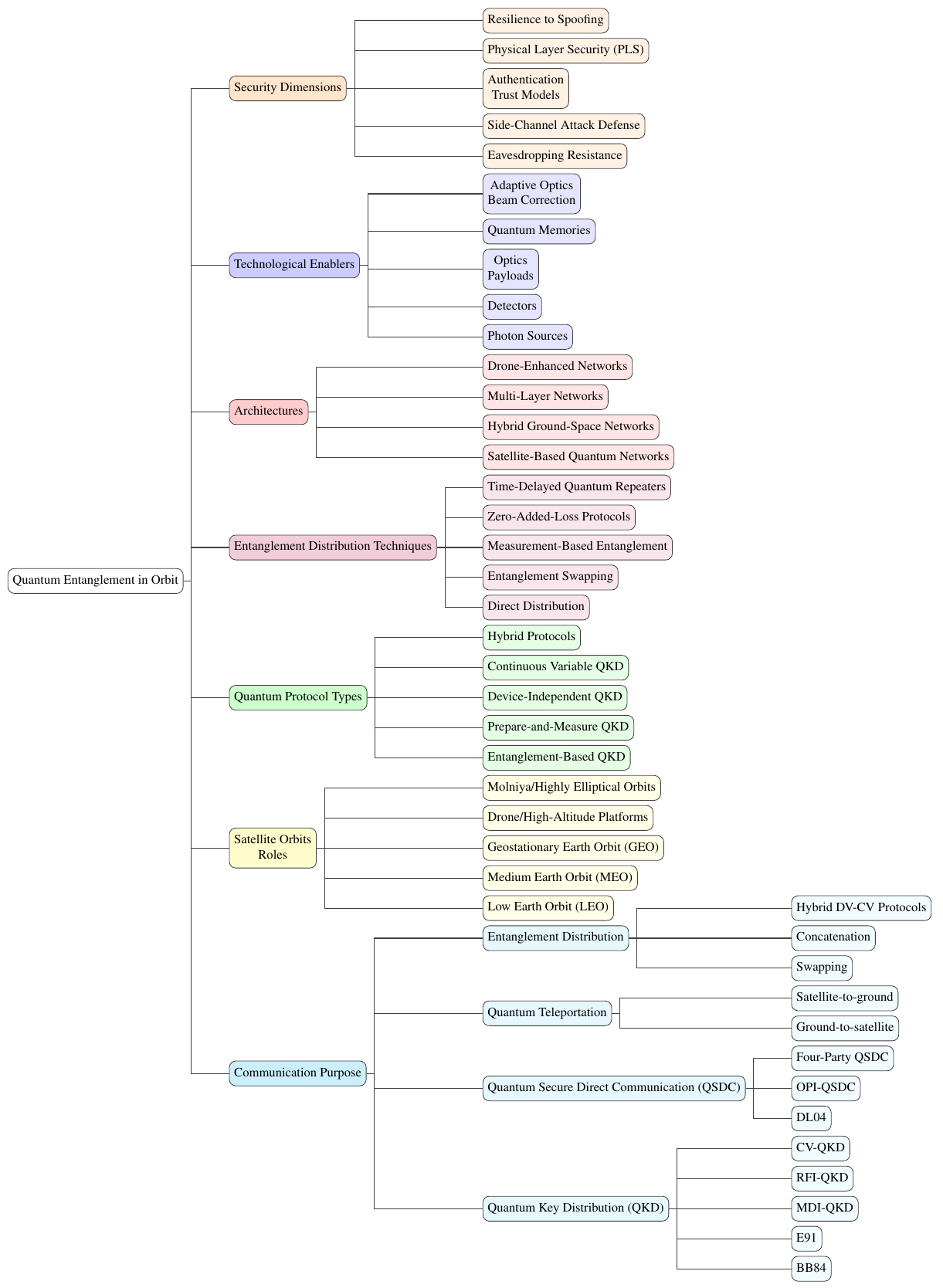}\\
\end{center}
\caption{Taxonomy of quantum entanglement in orbit.}
\label{fig:taxonomy-QSC}
\end{figure*}

The taxonomy of quantum entanglement in orbit (Fig.~\ref{fig:taxonomy-QSC}) provides a comprehensive framework for classifying the layers and components of Quantum Satellite Communication (QSC). At the functional level, it identifies key purposes, including QKD, QSDC, quantum teleportation, and entanglement distribution. QKD protocols (BB84, E91, MDI-QKD) support secure key generation with varying trust and security levels, while QSDC protocols (DL04, OPI-QSDC, Four-Party QSDC) enable direct message transmission. Quantum teleportation is addressed in both ground-to-satellite and satellite-to-ground links, and entanglement distribution is expanded through techniques like swapping, concatenation, and DV–CV hybrid schemes.

The taxonomy also categorizes satellite roles and orbital regimes. LEO satellites provide low latency and high key rates but limited access. MEO offers a balance between coverage and latency, while GEO ensures continuous links but with increased losses. Specialized orbits, such as Molniya, enhance high-latitude coverage and relay flexibility. From a protocol perspective, QKD approaches include entanglement-based (E91), prepare-and-measure (BB84), device-independent (DI-QKD, MDI-QKD), continuous-variable (CV QKD), and hybrid DV–CV models. This diversity reflects trade-offs in implementation complexity, performance, and resilience to side-channel attacks.

Supporting layers address entanglement distribution techniques, network architectures, and enabling technologies. Methods include direct links, relay-based swapping, and photon multiplexing. Architectures range from hierarchical networks to hybrid fiber–satellite systems (e.g., QuESat). Key technologies include photon sources (SPDC, quantum dots), advanced detectors (Si-SPADs, SNSPDs), and adaptive optics for atmospheric compensation. Finally, the taxonomy highlights security dimensions, detailing countermeasures against eavesdropping, spoofing, and side-channel attacks through principles like quantum no-cloning, decoy-state protocols, and entanglement-based authentication. This structured taxonomy not only captures the state-of-the-art in QSC but also provides a lens for evaluating future advancements and standardization efforts.

\begin{table*}[ht!]
\centering
\caption{Architectures of QSC}
\label{tab:qsc_arch}
\begin{tabular}{p{4.5cm}p{4cm}p{4cm}p{4cm}}
\toprule
\textbf{Architecture Type} & \textbf{Description} & \textbf{Key Advantages} & \textbf{Limitations / Challenges}\\
\midrule
\textbf{LEO Satellite Constellations} & Satellites at 200–2000 km for QKD \& entanglement distribution & Low latency, global coverage & Limited flyover time; frequent handovers\\
\textbf{MEO/GEO Architectures} & Mid/High orbit systems enabling broader coverage but higher latency & Longer flyover time, large footprint & Increased diffraction loss; lower SKR\\
\textbf{Hybrid Ground-Satellite Networks} & Integration of fiber-optic networks with satellite QKD channels & Extends range, combines terrestrial and space benefits & Requires synchronization and interoperability\\
\textbf{Hierarchical Constellations} & LEO for distribution, GEO for control/coordination (e.g., QuEsat architecture) & Global scalability; flexible control & Increased complexity, resource coordination \\
\textbf{Drone/High-Altitude Relays} & Intermediate airborne relays between satellite and ground & Lower cost, flexible, close-range link reliability & Limited range, weather dependence \\
\bottomrule
\end{tabular}
\end{table*}
\section{Current Architectures}
\label{Sec_arch}
The current architectures of QSC are delineated in Table~\ref{tab:qsc_arch}.

\subsection{Low Earth Orbit (LEO) Satellite Constellations}
\begin{figure}[h!]
    \centering
    \includegraphics[width=0.98\linewidth]{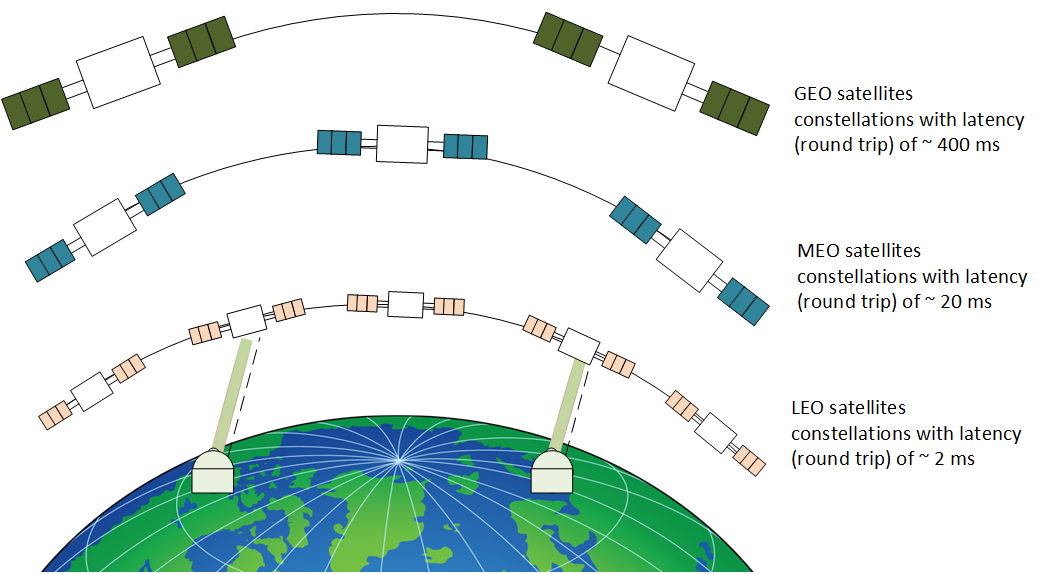}
    \caption{Satellite constellations architecture of GEO, MEO, and LEO}
    \label{leo}
\end{figure}

LEO satellites are pivotal in advancing global quantum networks due to their advantageous positioning at altitudes between 200 and 2,000 km, which offers broad coverage with reduced atmospheric interference and channel loss \cite{Wang2020AGT}. Unlike fiber-based quantum repeaters that face exponential absorption losses over long distances, LEO satellites primarily experience quadratic diffraction losses, enabling quantum communication well beyond the 2,000 km limit of terrestrial networks \cite{Goswami2023SatelliteRelayedGQ}. Besides quantum communication, LEO systems provide lower latency, higher data rates, and enhanced real-time capabilities compared to geostationary satellites \cite{Li2023PerformanceAO}. Their proximity and rapid orbital motion also enhance positioning and navigation accuracy. Fig.~\ref{leo} illustrates the architectural differences among GEO, MEO, and LEO constellations.

Examples of operational LEO constellations include QEYSSat, a Canadian mission focusing on Earth observation and communication, featuring LiDAR limb sounders for atmospheric and cloud measurements, particularly over polar regions \cite{Hoffmann2012ANS}. Similarly, China’s QUESS constellation targets quantum communication and navigation, utilizing QKD systems for secure links between satellites and ground stations, emphasizing coverage in the Asia-Pacific region \cite{Vergoossen2019SatelliteCF}.

\subsection{Satellite-Based QKD}
The Micius satellite is an LEO satellite that has served as a pioneering platform for a range of quantum experiments, including QKD, quantum teleportation, and entanglement distribution. Notably, it features a double-decker structural design, incorporating two optical transmitters, a space-based entangled photon source, an experimental control processor, and two acquisition, pointing, and tracking control modules \cite{Lu2022MiciusQE}. With a diameter of 2.5 meters and a mass of 880 kilograms, the Micius satellite significantly outperforms earlier systems in terms of transmission capabilities. Its transmission power has been reported to be ten times greater than that of conventional QKD satellite platforms, which enables downlink quantum communication over distances of up to 1,200 km \cite{Liao2018SatelliteRelayedIQ}. In practical deployments, the Micius QKD system has successfully established a quantum-secured communication link between the Zvenigorod and Nanshan ground stations, achieving a sifted key of 2.5 megabits and a final secure key of 310 kilobits \cite{Khmelev2023EurasianscaleES}.

\subsection{Hybrid Quantum-Classical Networks}
Satellites are increasingly being integrated with terrestrial fiber optic infrastructure to form hybrid quantum–classical communication systems, a development that has the potential to transform the landscape of global quantum communication \cite{Haldar2022TowardsGT}. This integration facilitates the establishment of global-scale quantum networks, with satellites playing a vital role in the distribution of entangled photons between ground stations.
A prominent approach involves the use of LEO satellites to deliver entanglement directly between designated terrestrial stations via quantum optical downlink channels \cite{Chang2023EntanglementDI}. In such architectures, GEO satellites can act as coordination nodes, tasked with overseeing and managing LEO satellite operations \cite{Conti2024SatelliteTerrestrialQN}. This forms a three-layered architecture that comprises GEO satellites, LEO satellites, and terrestrial ground stations. Within this framework, LEO satellites handle entanglement distribution, while GEO satellites facilitate coordination, thus offering broad coverage and helping to overcome the distance and infrastructure limitations of purely terrestrial fiber networks. Figure~\ref{leo} presents the satellite constellation architectures in which it highlights the primary distinctions between the GEO, MEO, and LEO orbits, particularly with respect to their respective distances from Earth.

The integration of satellite and terrestrial systems also enables the deployment of quantum communication protocols across extended distances \cite{Troupe2022QuantumCS}. In particular, the Micius satellite has already demonstrated the feasibility and effectiveness of such hybrid quantum communication channels on a scale \cite{Haldar2022TowardsGT}. More broadly, this integrated approach represents a promising path for the realization of large-scale robust quantum networks, offering global coverage and long-distance secure communication capabilities \cite{Djordjevic2020OnGQ}.

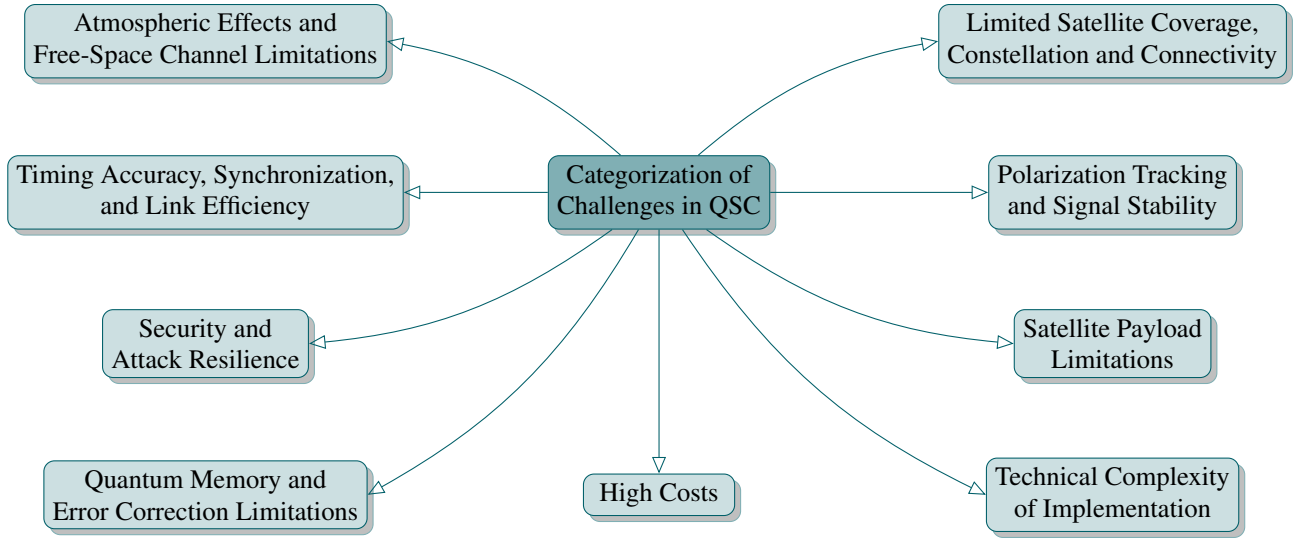
\begin{figure*}[ht!]
    \definecolor{colorgray}{HTML}{414141}
\definecolor{colorgreen}{HTML}{005F69}

\makeatletter
\tikzset{
    non-concept/.style={
        rectangle,
        execute at begin node=\footnotesize,
        text width=9em,
    },
    cncc east/.style={% concept-non-concept-connection
                      % where the non-concept is east of the concept
        out=0,
        in=180,
        to path={
            \pgfextra{
                \edef\tikztostart{\tikztostart.east}
                \edef\tikztotargetB{\tikztotarget.south east}
                \edef\tikztotarget{\tikztotarget.south west}
            }
            \tikz@to@curve@path% needs \makeatletter and \makeatother
            -- (\tikztotargetB)
        },
        draw=colorgreen, line width=0.3mm
    },
    cncc west/.style={% concept-non-concept-connection
                      % where the non-concept is west of the concept
        out=180,
        in=0,
        to path={
            \pgfextra{
                \edef\tikztostart{\tikztostart.west}
                \edef\tikztotargetB{\tikztotarget.south west}
                \edef\tikztotarget{\tikztotarget.south east}
            }
            \tikz@to@curve@path% needs \makeatletter and \makeatother
            -- (\tikztotargetB)
        },
        draw=colorgreen, line width=0.3mm % Change line style to dashed
    }
}
    \centering
    \begin{tikzpicture}[
        node distance=1.1cm,
        every node/.style={draw, rectangle, rounded corners, minimum width=2cm, fill=colorgreen!20, draw=colorgreen, drop shadow}, 
        every path/.style={draw=colorgreen, ->, >=open triangle 45}]
    
        % Nodes
        \node (solutions) [align=center, fill=colorgreen!50] {\baselineskip=12pt Categorization of \\ Challenges in QSC};
        \node (alt_train) [align=center, left of=solutions, node distance=6cm] {Timing Accuracy, Synchronization, \\ and Link Efficiency};
        \node (learn_stoch) [align=center, above of=alt_train, node distance=2cm] {Atmospheric Effects and \\ Free-Space Channel Limitations};
        \node (two_phase) [align=center, below of=alt_train, node distance=2cm] {Security and\\ Attack Resilience};
        \node (rl_base) [align=center, below of=two_phase, node distance=2cm] {Quantum Memory and \\ Error Correction Limitations};
        \node (err_mem) [align=center, below of=solutions, node distance=4cm] {High Costs};
    
        \node (genetic_algos) [align=center,right of=solutions, node distance=6cm] {Polarization Tracking\\ and Signal Stability};
        \node (grad_free) [align=center, above of=genetic_algos, node distance=2cm] {Limited Satellite Coverage, \\ Constellation and Connectivity};
        \node (diff_dsp) [align=center,below of=genetic_algos, node distance=2cm] {Satellite Payload\\ Limitations};
        \node (stoch_pert) [align=center,below of=diff_dsp, node distance=2cm] {Technical Complexity \\of Implementation};
    
        % Connections
        \draw (solutions) to (alt_train);
        \draw (solutions) to[bend left=-15](learn_stoch.east);
        \draw (solutions) to[bend left=15] (two_phase.east);
        \draw (solutions) to[bend left=15] (rl_base.east);
        \draw (solutions) to[bend left=15](grad_free.west);
        \draw (solutions) to (err_mem);
    
        \draw (solutions) to (genetic_algos);
        \draw (solutions) to[bend right=15] (diff_dsp.west);
        \draw (solutions) to[bend right=15] (stoch_pert.west);

    \end{tikzpicture}
    \caption{Categorization of QSC challenges}
    \label{fig_challenges}
\end{figure*}

\section{QSC Challenges}
\label{challenge}
QSC presents numerous challenges, ranging from technical difficulties in photon transmission to atmospheric interference and security concerns. Fig.~\ref{fig_challenges} provides a comprehensive overview of the primary challenges, which are detailed below.

\subsection{Atmospheric Effects and Free-Space Channel Limitations}
Fig.~\ref{fig_challenge_1} presents the major challenges and current solutions for atmospheric effects in QSC. As quantum signals traverse the atmosphere, they encounter various forms of impairments that degrade signal fidelity and reduce key generation rates. Atmospheric conditions significantly affect quantum communication signals. Factors such as scattering, absorption, diffraction, and turbulence introduce losses that degrade the quality of the QKD. These \textbf{atmospheric losses} degrade quantum signals and lower the efficiency of QKD \cite{yan2009atmospheric, Chakraborty2024AHN}. 

Diffraction loss is one of the main challenges in QSC, which results from the atmosphere of the Earth. This loss arises due to the finite aperture size of telescopes and the beam waist, which causes the quantum signal to spread and lose intensity during transmission \cite{Chakraborty2024AHN}. Such effects are particularly pronounced for high-orbit satellites, where the increased distance exacerbates beam divergence and significantly reduces transmission efficiency \cite{calderaro2018towards}. To mitigate diffraction loss, researchers commonly employ larger aperture telescopes and shorter photon wavelengths, both of which help to concentrate the beam and enhance signal strength. 

Atmospheric turbulence represents another substantial obstacle for reliable quantum communication, which causes \textbf{beam wandering and wavefront distortion}, especially in slant path configurations, causing signal broadening and loss \cite{chou2023satellite}. To address this, Chakraborty et al. \cite{Chakraborty2024AHN} have developed probability density functions to describe transmittance along slant propagation paths, proposed analytical models for quantum channels affected by turbulence, and carried out experimental characterization of long-distance atmospheric free-space optical (FSO) links. Typically, this type of loss can be mitigated by using \textbf{adaptive optical systems} and optimal ground station selection. These systems can compensate for wavefront distortion in real-time, especially in downlink configurations where correction is more practical \cite{sidhu2021advances}. \textbf{AI-based turbulence} can be used to predict atmospheric turbulence conditions and enable proactive adjustment such as TAROCCO \cite{jaouni2024predicting}

The nature of the atmospheric channel can have a large impact on the QKD performance. For example, in some circumstances, the QKD relative performance of Gaussian and non-Gaussian states can be reverse of that found in fixed-attenuation channels \cite{Hosseinidehaj2016CVQKDWG}. To address these challenges, researchers have proposed various approaches, including the use of \textbf{Gaussian-entangled states} in fading channels, \textbf{uncorrelated fading channels} \cite{Hosseinidehaj2016CVQKDWG}, and the development of probability density functions of transmittance for slant propagation paths \cite{Chakraborty2024AHN}. 

\begin{figure}[ht!]
    \definecolor{colorgreen}{HTML}{005F69}
\definecolor{leo-blue}{RGB}{64,115,158}
\definecolor{geo-orange}{RGB}{218,124,48}

\tikzset{box/.style={rectangle, draw=black, rounded corners=2pt, thick},
    param/.style={box, font=\normalsize, minimum width=4cm, minimum height=0.6cm, text width=7.5cm, align=left, drop shadow},
    header/.style={box, font=\Large, text=black, minimum width=5cm, minimum height=0.8cm, drop shadow},
    icon/.style={circle, draw, thin, minimum size=6mm}}

\centering
\resizebox{1.0\columnwidth}{!}{
\begin{tikzpicture}[y=-0.8cm]

% Column headers
\node[header, fill=colorgreen!30] at (0,-0.2) {Atmospheric Effects and Free-Space Channel Limitations};
\node[header, text=white, fill=leo-blue] at (4cm,1) {Solutions};
\node[header, text=white, fill=geo-orange] at (-4cm,1) {Challenges};

% challenges
\node[param, fill=geo-orange!30] at (-4,2.1) {\textbf{Atmospheric losses}: due to absorption, scattering, diffraction, and turbulence};
\node[param, fill=geo-orange!30] at (-4,3.3) {\textbf{Beam wandering and wavefront distortion}: especially significant during slant path propagation};
\node[param, fill=geo-orange!30] at (-4,4.45) {\textbf{Point-ahead angle limitations:} in uplink configurations};
\node[param, fill=geo-orange!30] at (-4,5.6) {\textbf{Wavelength-dependent attenuation}: variable impact across C-band, Si-band};
\node[param, fill=geo-orange!30] at (-4,6.8) {\textbf{Beam divergence in higher orbits}: degrades signal strength at longer distances};

% Solutions
\node[param, fill=leo-blue!30] at (4,2.05) {\textbf{Adaptive optics systems} to counter wavefront distortion};
\node[param, fill=leo-blue!30] at (4,3.17) {\textbf{AI-based turbulence forecasting} TAROCCO RNN model};
\node[param, fill=leo-blue!30] at (4,4.34) {\textbf{Gaussian-entangled states} in fading channels, and uncorrelated fading channels};
\node[param, fill=leo-blue!30] at (4,5.55) {\textbf{Optimized wavelength selection} based on atmospheric transparency};
\node[param, fill=leo-blue!30] at (4,6.75) {\textbf{Tailored optics} autorotatable half-wave plates and polarization-preserving films};

\end{tikzpicture}
}
    \caption{Atmospheric effects and free-space channel limitations}
    \label{fig_challenge_1}
\end{figure}

Uplink communication paths in quantum satellite systems often require the use of a \textbf{point-ahead angle} to compensate for satellite motion. This angle typically exceeds the isoplanatic angle, introducing complexity into adaptive optics correction mechanisms and reducing the accuracy of beam alignment \cite{sidhu2021advances}. \textbf{Wavelength-dependent attenuation} is another issue in which atmospheric conditions affect different wavelength bands to varying degrees. As a result, the selection of operational wavelengths must be tailored to the specific environmental conditions in which the system is deployed \cite{Hosseinidehaj2016CVQKDWG}. For satellites operating in MEO or GEO, \textbf{beam divergence} becomes increasingly significant due to the greater transmission distance. This divergence leads to reduced link efficiency, further complicating the establishment of high-fidelity quantum channels over long ranges \cite{calderaro2018towards}. \textbf{Optimized wavelength selection} by using wavelength bands tailored to specific atmospheric conditions (e.g., C-band vs. Si-band) improves transmission fidelity \cite{Hosseinidehaj2016CVQKDWG}. One of the solution for signal stability over long distance is tailored optics by using Polarization-preserving films and autorotatable half-wave plates \cite{Lu2022MiciusQE}.

\subsection{Timing Accuracy, Synchronization, and Link Efficiency}
QSC systems rely heavily on precise timing and synchronization to securely exchange quantum keys between satellites and ground stations. However, achieving the required level of timing accuracy remains a major challenge, as current satellite technologies often fall short of delivering sub-nanosecond precision \cite{agnesi2019sub}. While classical clock synchronization techniques, including those employing optical frequency combs, have demonstrated femtosecond-level precision under laboratory conditions \cite{Gosalia2024ClassicalAQ}, these methods typically lack scalability and are not well suited to the constraints of space-based applications. As an alternative, quantum clock synchronization (QCS) has been proposed, which utilities entangled photon pairs to synchronize clocks across distant locations with sub-nanosecond accuracy. Recent proposals suggest the potential for implementing QCS on a global scale by deploying a quantum-assisted master clock in orbit. As a result, it enables continuous global synchronization with unprecedented precision \cite{Ducoing2023AQM}.

\begin{figure}[ht!]
    \definecolor{colorbraun}{HTML}{b67d77}
\definecolor{colorlila}{HTML}{7d77b6}
\definecolor{colorgreen}{HTML}{005F69}

\tikzset{box/.style={rectangle, draw=black, rounded corners=2pt, thick},
    param/.style={box, font=\normalsize, minimum width=4cm, minimum height=0.6cm, text width=7.5cm, align=left, drop shadow},
    header/.style={box, font=\Large, text=black, minimum width=5cm, minimum height=0.8cm, drop shadow},
    icon/.style={circle, draw, thin, minimum size=6mm}}

\centering
\resizebox{1.0\columnwidth}{!}{
\begin{tikzpicture}[y=-0.8cm]

% Column headers
\node[header, fill=colorgreen!30] at (0,-0.2) {Timing Accuracy, Synchronization, and Link Efficiency};
\node[header, text=white, fill=colorlila] at (4cm,1) {Solutions};
\node[header, text=white, fill=colorbraun] at (-4cm,1) {Challenges};

% Challenges
\node[param, fill=colorbraun!30] at (-4,2.1) {\textbf{Sub-nanosecond timing precision} is required for quantum key exchanges};
\node[param, fill=colorbraun!30] at (-4,3.26) {\textbf{Clock skew and satellite-to-ground synchronization}};
\node[param, fill=colorbraun!30] at (-4,4.2) {\textbf{Inter-satellite link latency and instability}};
\node[param, fill=colorbraun!30] at (-4,5.14) {\textbf{Photon loss and finite-size effects} affecting QKD rate};
\node[param, fill=colorbraun!30] at (-4,6.32) {\textbf{Beam alignment accuracy} required for stable entangled photon delivery};

% Solutions
\node[param, fill=colorlila!30] at (4,2.11) {\textbf{Quantum clock synchronization (QCS)} using entangled photons and \textbf{master clock architectures}};
\node[param, fill=colorlila!30] at (4,3.1) {\textbf{Quantum frequency combs} for scalable precision};
\node[param, fill=colorlila!30] at (4,4.05) {\textbf{Bright, space-hardened entangled photon sources}};
\node[param, fill=colorlila!30] at (4,5.0) {\textbf{Sun-synchronous orbits and CV-QKD}};
\node[param, fill=colorlila!30] at (4,5.98) {\textbf{Enhanced tracking systems} for maintaining beam alignment};

\end{tikzpicture}
}
    \caption{Timing Accuracy, Synchronization, and Link Efficiency}
    \label{fig_challenge_2}
\end{figure}

As shown in Fig.~\ref{fig_challenge_2}, current challenges regarding timing precision and synchronization complications include:

\begin{itemize}
    \item \textbf{Sub-nanosecond timing precision:} It is essential for high-fidelity QKD, but difficult to achieve in satellite systems \cite{agnesi2019sub}.
    \item \textbf{Clock frequency skew:} The skew in clock frequency is a significant challenge in QCS, as it can lead to synchronization errors and degrade entanglement-based performance. To address this challenge, several techniques have been proposed, including the use of compensation with an effective frequency skew of less than 48 ns/s and the use of correlated photons \cite{Spiess2021ClockSW}.
    \item \textbf{Satellite-to-ground synchronization:} It is a challenging task due to the high-loss free-space channel and the limited stability of satellite clocks \cite{Haldar2022TowardsGT}. To address this challenge, several techniques have been proposed, including the use of binary-phase-modulated continuous-wave laser and the use of a predefined random sequence for timing analysis \cite{Zhang2021TimingAS}.
    \item \textbf{Inter-satellite links:} Inter-satellite links pose another challenge due to the limited communication capacity and the high-loss free-space channel. To address this challenge, several techniques have been proposed, including the use of quantum frequency combs and the use of entangled photons \cite{Ducoing2023AQM}.
    \item \textbf{Photon loss and finite-size effects:} this can decrease secret key rates and reduce the reliability of QKD links \cite{pirandola2021satellite}.
    \item \textbf{Beam Alignment Accuracy}: It is due to the rotation and movement of satellites that can introduce polarization fluctuations. Several solutions have been proposed to address this challenge, such as reference-frame-independent (RFI) protocols \cite{laing2010reference}, augmenting quantum communications protocols \cite{higgins2018practical},  polarization frame alignment protocols \cite{higgins2018practical} and correlated side-channel measurements \cite{madani2025quantum}.
\end{itemize}

As summarized in Fig.~\ref{fig_challenge_2}, the current solutions for timing accuracy, synchronization can be categorized into the following:
\begin{itemize}
    \item \textbf{Quantum clock synchronization:} It uses entangled photons for sub-nanosecond synchronization across nodes \cite{Ducoing2023AQM}. Entangled photons have been shown to be a valuable resource for synchronization with minimal hardware overhead \cite{Spiess2021ClockSW}. The use of entangled photons in QCS provides a simple way to enhance the timing resolution in distributed quantum information processing tasks.
    \item \textbf{Quantum-assisted master clock in the sky:} A quantum-assisted master clock in the sky has been proposed as a method to extend QCS to a global scale \cite{Ducoing2023AQM}. This approach could provide continuous global synchronization at the sub-nanosecond level using a quantum-assisted master clock in the sky.    
    \item \textbf{Quantum frequency combs:} They have the potential to provide precision-to-resource advantages that go beyond the capabilities of classical frequency combs \cite{Gosalia2024ClassicalAQ}. In resource-constrained environments such as satellite networks, quantum frequency combs could be a better solution than classical frequency combs.
    \item \textbf{Bright entangled photon sources and CV-QKD protocols:}  It can enhance link efficiency and timing robustness \cite{pirandola2021satellite}.
    \item \textbf{Sun-synchronous orbits and orbit selection:} It can improve QKD efficiency by optimizing satellite pass frequency and coverage \cite{pirandola2021satellite}.    
    \item Compensation with an effective frequency skew: Compensation with an effective frequency skew less than 48 ns/s has been shown to be effective in reducing synchronization errors \cite{Spiess2021ClockSW}. This technique could be used to improve the performance of QCS in high-loss free-space channels.
\end{itemize}

\subsection{Security and Attack Resilience}
\begin{figure}[ht!]
    \definecolor{colorbraun}{HTML}{ad3d5a}
\definecolor{colorlila}{HTML}{3d92ad}
\definecolor{colorgreen}{HTML}{005F69}
\tikzset{box/.style={rectangle, draw=black, rounded corners=2pt, thick},
    param/.style={box, font=\normalsize, minimum width=4cm, minimum height=0.6cm, text width=7.5cm, align=left, drop shadow},
    header/.style={box, font=\Large, text=black, minimum width=5cm, minimum height=0.8cm, drop shadow},
    icon/.style={circle, draw, thin, minimum size=6mm}}

\centering
\resizebox{1.0\columnwidth}{!}{
\begin{tikzpicture}[y=-0.8cm]

% Column headers
\node[header, fill=colorgreen!30] at (0,-0.2) {Security and Attack Resilience};
\node[header, text=white, fill=colorlila] at (4cm,1) {Solutions};
\node[header, text=white, fill=colorbraun] at (-4cm,1) {Challenges};

% Challenges
\node[param, fill=colorbraun!30] at (-4,1.88) {\textbf{Side-channel and physical-layer attacks}};
\node[param, fill=colorbraun!30] at (-4,2.87) {\textbf{Prepare-and-measure QKD vulnerabilities} to spoofing and jamming};
\node[param, fill=colorbraun!30] at (-4,4.05) {\textbf{Dependency on trusted nodes} in current architectures};
\node[param, fill=colorbraun!30] at (-4,5.0) {\textbf{Vulnerabilities to laser-based ground attacks}};

% Solutions
\node[param, fill=colorlila!30] at (4,2.11) {\textbf{Measurement-device-independent QKD (MDI-QKD)} to remove detector trust assumptions.};
\node[param, fill=colorlila!30] at (4,3.33) {\textbf{Hardware-level countermeasures} for physical-layer threats.};
\node[param, fill=colorlila!30] at (4,4.5) {\textbf{End-to-end secure QKD protocols} avoiding intermediate trust.};
\node[param, fill=colorlila!30] at (4,5.67) {\textbf{FSO technologies} combined with QKD for resilient and secure high-throughput links.};

\end{tikzpicture}
}
    \caption{Security and attack resilience}
    \label{fig_challenge_3}
\end{figure}
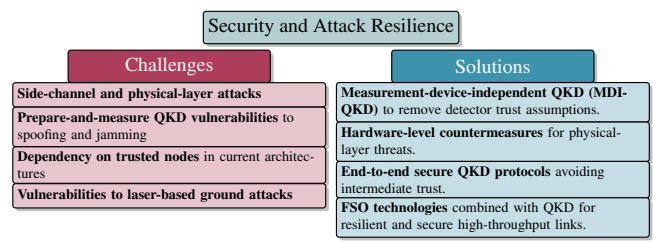

QSC systems remain vulnerable to a range of security threats, including eavesdropping, jamming, and spoofing attacks \cite{Kareem2024CyberTL}. In particular, a sufficiently well-resourced adversary may be capable of transmitting unauthorized commands to a satellite, potentially gaining access to sensitive data or assuming control of satellite operations with the potential to inflict serious disruption or damage \cite{AlHraishawi2021BroadbandNS}. Additionally, satellites subjected to denial-of-service (DoS) attacks may expend considerable computational resources and processing time handling spurious or malicious messages. This not only depletes system performance but also degrades the quality of service available to legitimate users. Fig.~\ref{fig_challenge_3} presents the major challenges and current solutions for security and attack resilience in QSC.

To address these security challenges, various security mechanisms have been proposed, including the use of QKD. QKD is a secure communication method that uses quantum mechanics to encode and decode messages \cite{Rani2025CombinedQA}. This method is resistant to eavesdropping and can provide unconditional security. Although QKD is theoretically unbreakable, practical implementations still have vulnerabilities, such as \textbf{side-channel attacks}, that require robust countermeasures \cite{chou2023satellite}. Implementing \textbf{measurement-device-independent QKD} can eliminate security vulnerabilities associated with detector-side attacks without requiring a trusted satellite. Moreover, \textbf{hardware-based countermeasures} against side-channel attacks can enhance the security of QSC \cite{pirandola2021satellite}. 
In addition, QKD has its own set of challenges, such as traditional \textbf{prepare-and-measure QKD protocols} require trusted satellite nodes, introducing potential security vulnerabilities \cite{orsucci2024assessment}. Implementing fully device-independent protocols in satellite environments remains a challenge due to efficiency and practical constraints, such as developing \textbf{end-to-end security protocols} that do not rely on trusted intermediary nodes can provide stronger security guarantees \cite{pirandola2021satellite}.

Beyond QKD, several alternative security mechanisms have been proposed for the robustness of QSC systems. One such approach involves the adoption of \textbf{free-space optical (FSO) communication technology}, which offers a wide bandwidth and high data rates, which presents a compelling alternative to traditional radio frequency (RF) intersatellite links \cite{AlHraishawi2021BroadbandNS}. In addition, physical-layer security techniques have been explored within the context of satellite communication \cite{Burleigh2019FromCT}. These methods exploit the inherent physical characteristics of the communication channel to provide security, for example, by using entangled photons to ensure unconditional security in information exchange between remote parties. However, despite these advances, satellite-based quantum communication remains vulnerable to a range of attack vectors. Notably, \textbf{ground-based laser attacks} have been identified as a significant threat, capable of disrupting QKD links over distances of up to 1,000 km \cite{Gozzard2021VulnerabilityOS}.

\subsection{Quantum Memory and Error Correction Limitations}

A key limitation in the development of QSC systems is the current state of quantum memory technologies. One of the main challenges is the \textbf{short coherence time} exhibited by existing quantum memories, which significantly limits their applicability in quantum repeater architectures for the long-distance entanglement distribution \cite{gundougan2021proposal}. In addition, the harsh space environment characterized by radiation exposure and extreme temperature fluctuations poses further difficulties for the stable operation of quantum memory systems in orbit. The challenge is further compounded by integration issues, as current memory devices often struggle to meet the \textbf{size, weight, and power} (SWaP) constraints typical of satellite platforms \cite{Haldar2022TowardsGT}. Moreover, many memory-assisted quantum communication protocols are not yet fully optimized for satellite-specific operating conditions, which limits their efficiency and scalability in practice \cite{gundougan2021proposal}.

\begin{figure}[ht!]
    \definecolor{colorbraun}{HTML}{ad3d92}
\definecolor{colorlila}{HTML}{3dad90}
\definecolor{colorgreen}{HTML}{005F69}
\tikzset{box/.style={rectangle, draw=black, rounded corners=2pt, thick},
    param/.style={box, font=\normalsize, minimum width=4cm, minimum height=0.6cm, text width=7.5cm, align=left, drop shadow},
    header/.style={box, font=\Large, text=black, minimum width=5cm, minimum height=0.8cm, drop shadow},
    icon/.style={circle, draw, thin, minimum size=6mm}}

\centering
\resizebox{1.0\columnwidth}{!}{
\begin{tikzpicture}[y=-0.8cm]

% Column headers
\node[header, fill=colorgreen!30] at (0,-0.2) {Quantum Memory and Error Correction Limitations};
\node[header, text=white, fill=colorlila] at (4cm,1) {Solutions};
\node[header, text=white, fill=colorbraun] at (-4cm,1) {Challenges};

% Challenges
\node[param, fill=colorbraun!30] at (-4,1.89) {\textbf{Short coherence times} in quantum memories.};
\node[param, fill=colorbraun!30] at (-4,2.88) {\textbf{Radiation sensitivity and size/power limits} in spaceborne memory.};
\node[param, fill=colorbraun!30] at (-4,3.87) {\textbf{No standardized QEC for satellite links}};
\node[param, fill=colorbraun!30] at (-4,4.81) {\textbf{Resource-heavy error correction codes} incompatible with satellite constraints.};

% Solutions
\node[param, fill=colorlila!30] at (4,2.09) {\textbf{Dual-memory satellite designs} with varied coherence times.};
\node[param, fill=colorlila!30] at (4,3.27) {\textbf{Room-temperature quantum memories} for space operation.};
\node[param, fill=colorlila!30] at (4,4.46) {\textbf{Quantum memories with error correction techniques}};
\node[param, fill=colorlila!30] at (4,5.43) {\textbf{Compact, AI-enhanced QEC protocols}};

\end{tikzpicture}
}
    \caption{Quantum memory and error correction limitations}
    \label{fig_challenge_4}
\end{figure}
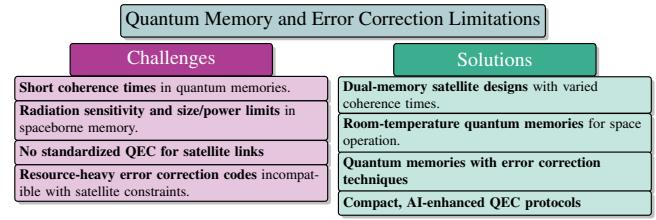

To address these research gaps, several potential solutions have been proposed. Fig.~\ref{fig_challenge_4} shows these challenges and current solutions. One innovative concept involves the deployment of \textbf{dual-memory satellite systems}, which incorporate quantum memories with both long and short coherence times. This configuration offers increased flexibility and efficiency for global-scale quantum networking \cite{gundogan2024time}. Another promising direction is the development of \textbf{room-temperature quantum memories}, which would significantly reduce both power consumption and system complexity, making them more suitable for spaceborne platforms \cite{gundougan2021proposal}. Furthermore, integration of \textbf{quantum memories with error correction techniques} has been suggested as a means of extending effective storage times and enhancing transmission fidelity \cite{terhal2015quantum, bravyi2024high, nakazato2022quantum}. Finally, AI can be used to optimize the design and implementation of QEC by dynamically adjusting error correction strategies based on real-time data \cite{ma2020optical}. This can lead to improved performance in QSC systems.

\subsection{Limited Satellite Coverage, Constellation Management and Connectivity}
\begin{figure}[ht!]
    \definecolor{colorbraun}{HTML}{796b6d}
\definecolor{colorlila}{HTML}{297360}
\definecolor{colorgreen}{HTML}{005F69}
\tikzset{box/.style={rectangle, draw=black, rounded corners=2pt, thick},
    param/.style={box, font=\normalsize, minimum width=4cm, minimum height=0.6cm, text width=7.5cm, align=left, drop shadow},
    header/.style={box, font=\Large, text=black, minimum width=5cm, minimum height=0.8cm, drop shadow},
    icon/.style={circle, draw, thin, minimum size=6mm}}

\centering
\resizebox{1.0\columnwidth}{!}{
\begin{tikzpicture}[y=-0.8cm]

% Column headers
\node[header, fill=colorgreen!30] at (0,-0.2) {Satellite Coverage, Constellation Management, and Connectivity};
\node[header, text=white, fill=colorlila] at (4cm,1) {Solutions};
\node[header, text=white, fill=colorbraun] at (-4cm,1) {Challenges};

% Challenges
\node[param, fill=colorbraun!30] at (-4,2.12) {\textbf{Limited ground station visibility} from fast-moving LEO satellites};
\node[param, fill=colorbraun!30] at (-4,3.33) {\textbf{Line-of-sight (LoS) constraints} restricting continuous connectivity};
\node[param, fill=colorbraun!30] at (-4,4.31) {\textbf{High channel loss from MEO/GEO orbits}};
\node[param, fill=colorbraun!30] at (-4,5.3) {\textbf{Ultra-dense LEO satellite constellations}, efficient beam coverage and interference};

% Solutions
\node[param, fill=colorlila!30] at (4,2.12) {\textbf{Movable antenna arrays} to optimize dynamic beam coverage.};
\node[param, fill=colorlila!30] at (4,3.34) {\textbf{Constellation-based QKD systems} for scalable global coverage};
\node[param, fill=colorlila!30] at (4,4.51) {\textbf{Trusted relay satellite networks} to extend key distances};
% \node[param, fill=colorlila!30] at (4,5.0) {\textbf{Time-delayed repeater protocols} with dual-memory satellites};

\end{tikzpicture}
}
    \caption{Satellite coverage, constellation management, and connectivity}
    \label{fig_challenge_5}
\end{figure}
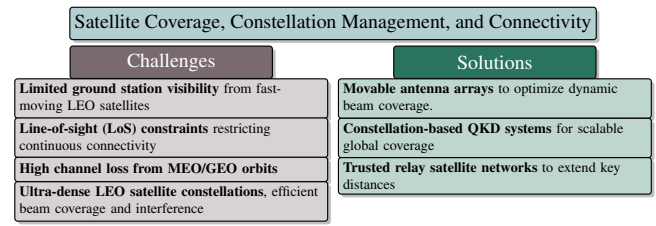

Satellite-based QKD can provide global coverage and overcome the limitations of terrestrial QKD, which is limited by fiber absorption and scales exponentially with distance \cite{Wang2022DeploymentSF}. However, satellite-based QKD also faces challenges due to limited satellite coverage and connectivity. The availability of quantum-secure communication is limited by the number of quantum-enabled satellites. Efficient global coverage requires a satellite network, which is costly and technologically demanding \cite{hughes2010satellite}. Fig.~\ref{fig_challenge_5} summarizes these challenges and potential solutions.

One of the primary challenges in QSC lies in the \textbf{limited coverage area of LEO satellites}, which are generally preferred due to their low altitude and correspondingly reduced channel loss \cite{Wang2022DeploymentSF}. However, the rapid orbital motion of LEO satellites significantly restricts the flyover duration for each ground station, making it difficult to maintain stable and continuous connections. In contrast, satellites in MEO and GEO offer wider coverage and longer visibility windows but suffer from increased channel attenuation and lower key generation rates. Another fundamental constraint is the need for \textbf{a line-of-sight (LoS)} link between satellites and ground stations, which limits the effective separation between ground stations to less than 1,000 km in the case of LEO systems \cite{Wang2022DeploymentSF}. This restriction, determined by satellite altitude, poses a significant obstacle to the deployment of a truly global QKD network. Moreover, the increasing use of \textbf{ultra-dense LEO satellite constellations} introduces additional complexity in terms of beam coverage management and interference mitigation \cite{Zhu2024DynamicBC}. Traditional directional antennas and fixed-position antenna arrays offer limited flexibility in beamforming, making it difficult to adapt to the dynamic coverage requirements of terrestrial users.

To address these challenges, several promising solutions have been proposed. One approach involves the use of \textbf{movable antenna arrays}, which offer greater flexibility in beam coverage and enhanced capabilities for interference mitigation \cite{Zhu2024DynamicBC}. These adaptive systems allow satellite constellations to better accommodate the dynamic spatial demands of terrestrial users. Another widely explored strategy is the deployment of \textbf{trusted relays}, such as intermediary satellites, which can theoretically extend QKD links indefinitely and facilitate global-scale coverage \cite{Wang2022DeploymentSF}. However, this approach relies on the availability of simultaneous LoS connections between the relay satellite and both participating ground stations, a requirement that presents practical difficulties.

\subsection{Polarization Tracking and Signal Stability}
\begin{figure}[ht!]
    \definecolor{colorbraun}{HTML}{e20222}
\definecolor{colorlila}{HTML}{0292e2}
\definecolor{colorgreen}{HTML}{005F69}
\tikzset{box/.style={rectangle, draw=black, rounded corners=2pt, thick},
    param/.style={box, font=\normalsize, minimum width=4cm, minimum height=0.6cm, text width=7.5cm, align=left, drop shadow},
    header/.style={box, font=\Large, text=black, minimum width=5cm, minimum height=0.8cm, drop shadow},
    icon/.style={circle, draw, thin, minimum size=6mm}}

\centering
\resizebox{1.0\columnwidth}{!}{
\begin{tikzpicture}[y=-0.8cm]

% Column headers
\node[header, fill=colorgreen!30] at (0,-0.2) {Polarization Tracking and Signal Stability};
\node[header, text=white, fill=colorlila] at (4cm,1) {Solutions};
\node[header, text=white, fill=colorbraun] at (-4cm,1) {Challenges};

% Challenges
\node[param, fill=colorbraun!30] at (-4,2.07) {\textbf{Polarization drift and misalignment} due to satellite movement};
\node[param, fill=colorbraun!30] at (-4,3.2) {\textbf{Signal degradation} from fluctuating environmental conditions};

% Solutions
\node[param, fill=colorlila!30] at (4,1.89) {\textbf{Active feedback polarization tracking systems}};
\node[param, fill=colorlila!30] at (4,2.84) {\textbf{Tomography-informed optimal measurement bases}};
\node[param, fill=colorlila!30] at (4,3.78) {\textbf{Tailored optical films}};
\node[param, fill=colorlila!30] at (4,4.77) {\textbf{Autorotatable optical elements} for real-time polarization correction};

\end{tikzpicture}
}
    \caption{Polarization tracking and signal stability}
    \label{fig_challenge_6}
\end{figure}
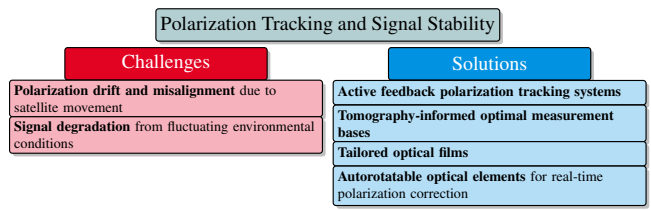

Polarization tracking is a critical element in QSC, as it ensures the proper alignment of the polarization states of photons transmitted between satellites and ground stations, as shown in Fig.~\ref{fig_challenge_6}. However, during long-distance transmission, the polarization state of light is inevitably subject to fluctuations, which results in signal instability and degradation in communication quality. These variations are further compounded by satellite motion and environmental disturbances, which necessitate the use of advanced tracking mechanisms \cite{wang2014polarization}.

Several approaches have been proposed to address this challenge. One widely studied method is the use of \textbf{active feedback-based control systems}, which dynamically adjust the polarization of the transmitter to align with that of the receiver. While effective, these systems are often resource intensive and may incur additional maintenance costs \cite{Toyoshima2011PolarizationBasisTS}. An alternative strategy involves the construction of \textbf{optimal measurement bases}, designed to minimize the QBER and maximize key generation rates, even in the presence of polarization fluctuations \cite{Chatterjee2022PolarizationCT}. This approach relies on quantum-state tomography of the received two-qubit state prior to each QKD session, using the reconstructed density matrix to determine the most effective measurement configuration. In addition to these methods, other techniques have been explored to enhance signal stability. For example, the application of \textbf{tailored optical films} has been shown to preserve high polarization extinction ratios within the optical system, thereby mitigating polarization drift \cite{Lu2022MiciusQE}. Furthermore, the integration of \textbf{autorotatable half-wave plates} offers a promising solution for maintaining polarization alignment dynamically during transmission.

\subsection{Satellite Payload Limitations}
\begin{figure}[ht!]
    \definecolor{colorbraun}{HTML}{ff003b}
\definecolor{colorlila}{HTML}{008b8b}
\definecolor{colorgreen}{HTML}{005F69}
\tikzset{box/.style={rectangle, draw=black, rounded corners=2pt, thick},
    param/.style={box, font=\normalsize, minimum width=4cm, minimum height=0.6cm, text width=7.5cm, align=left, drop shadow},
    header/.style={box, font=\Large, text=black, minimum width=5cm, minimum height=0.8cm, drop shadow},
    icon/.style={circle, draw, thin, minimum size=6mm}}

\centering
\resizebox{1.0\columnwidth}{!}{
\begin{tikzpicture}[y=-0.8cm]

% Column headers
\node[header, fill=colorgreen!30] at (0,-0.2) {Satellite Payload Limitations};
\node[header, text=white, fill=colorlila] at (4cm,1) {Solutions};
\node[header, text=white, fill=colorbraun] at (-4cm,1) {Challenges};

% Challenges
\node[param, fill=colorbraun!30] at (-4,2.12) {\textbf{limited volume and weight available} of the payload};
\node[param, fill=colorbraun!30] at (-4,3.11) {\textbf{Microgravity operational challenges}};
\node[param, fill=colorbraun!30] at (-4,4.1) {\textbf{Thermal management}, \textbf{radiation} requirements, and \textbf{micrometeoroid} impacts};
\node[param, fill=colorbraun!30] at (-4,5.08) {\textbf{Fault tolerance and long-duration reliability} };

% Solutions
\node[param, fill=colorlila!30] at (4,1.88) {\textbf{Quantum communication hub mission}};
\node[param, fill=colorlila!30] at (4,2.66) {\textbf{Microsatellite-based solutions}};
\node[param, fill=colorlila!30] at (4,3.6) {\textbf{Quantum Research Cubesat (QUARC) constellation}};
\node[param, fill=colorlila!30] at (4,4.54) {\textbf{Dual-purpose optical payloads}};

\end{tikzpicture}
}
    \caption{Satellite payload limitations}
    \label{fig_challenge_6_5}
\end{figure}
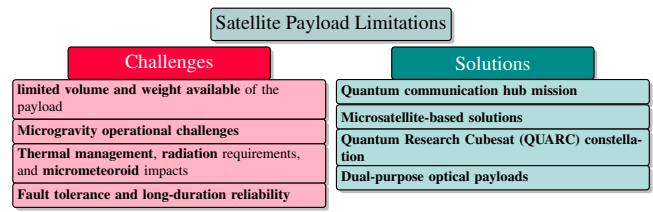

Satellite payload limitations pose a critical challenge in the design and operation of Quantum Satellite Communication (QSC) systems (Fig.~\ref{fig_challenge_6_5}). These constraints stem primarily from the restricted volume, weight, and power budgets inherent in satellite platforms, particularly in microsatellites. For instance, the Quantum Communication Hub operates within a shared 2U payload of a 12U platform, requiring highly compact and optimized internal designs \cite{Sagar2022DesignAT}. Similarly, limited power—defined by the satellite’s solar panels and energy storage—necessitates careful power management to support both data processing and secure transmission \cite{Mendes2023OpticalPD}.

Environmental factors in space introduce additional engineering complexities. Payloads must perform reliably under microgravity, radiation, thermal extremes, and potential micrometeoroid impacts \cite{Galetsky2022LEOSS}. As in-orbit maintenance is infeasible, systems must be engineered for high reliability and autonomy, validated through rigorous testing and fault-tolerant designs \cite{Li2024MicrosatellitebasedRQ}. Security is equally critical: payloads must safeguard sensitive quantum data through encryption, secure protocols, and tamper-resistant architectures \cite{Orsucci2024AssessmentOP}. Moreover, scalability and modularity are essential to accommodate future demand and technology upgrades \cite{Zhang2023EndtoendDF}.

To address these challenges, several strategies have emerged. Compact payload solutions, like the Quantum Communication Hub \cite{sagar2023design}, enable efficient Quantum Key Distribution (QKD) within stringent resource limits. Microsatellite-based platforms \cite{li2025microsatellite} provide a cost-effective balance between performance and scalability, supporting constellation architectures such as QUARC \cite{mazzarella2020quarc}. These 6U CubeSats operate flexibly as trusted or untrusted nodes, enabling uplink or downlink modes in constellation deployments. Another innovation is dual-purpose optical payloads \cite{mendes2024optical}, supporting both QKD and Quantum Key Pooling Communication (QKPC). Simulations show such systems, even in a compact 3U CubeSat, can achieve secret key rates up to 80 kHz with minimal QBER (~0.07\%) using simplified BB84 protocols. These advancements demonstrate that thoughtful design, modular scalability, and constellation-based strategies can effectively mitigate payload constraints, paving the way for robust and scalable quantum satellite networks.

\subsection{Technical Complexity of Implementation}
\begin{figure}[ht!]
    \definecolor{colorbraun}{HTML}{e25202}
\definecolor{colorlila}{HTML}{02af40}
\definecolor{colorgreen}{HTML}{005F69}
\tikzset{box/.style={rectangle, draw=black, rounded corners=2pt, thick},
    param/.style={box, font=\normalsize, minimum width=4cm, minimum height=0.6cm, text width=7.5cm, align=left, drop shadow},
    header/.style={box, font=\Large, text=black, minimum width=5cm, minimum height=0.8cm, drop shadow},
    icon/.style={circle, draw, thin, minimum size=6mm}}

\centering
\resizebox{1.0\columnwidth}{!}{
\begin{tikzpicture}[y=-0.8cm]

% Column headers
\node[header, fill=colorgreen!30] at (0,-0.2) {Technical Complexity of Implementation};
\node[header, text=white, fill=colorlila] at (4cm,1) {Solutions};
\node[header, text=white, fill=colorbraun] at (-4cm,1) {Challenges};

% Challenges
\node[param, fill=colorbraun!30] at (-4,2.08) {\textbf{Precise alignment and stabilization of laser beams}};
\node[param, fill=colorbraun!30] at (-4,3.26) {\textbf{Size, Weight, and Power (SWaP) constraints}  on satellite payloads};
\node[param, fill=colorbraun!30] at (-4,4.24) {\textbf{Sensitivity of quantum states}};

% Solutions
\node[param, fill=colorlila!30] at (4,2.11) {\textbf{Miniaturized quantum components} with lower power profiles.};
\node[param, fill=colorlila!30] at (4,3.1) {\textbf{Detected laser beacon systems }};
\node[param, fill=colorlila!30] at (4,4.05) {\textbf{Modular satellite payload architectures} for scalability.};

\end{tikzpicture}
}
    \caption{Technical complexity of implementation}
    \label{fig_challenge_7}
\end{figure}
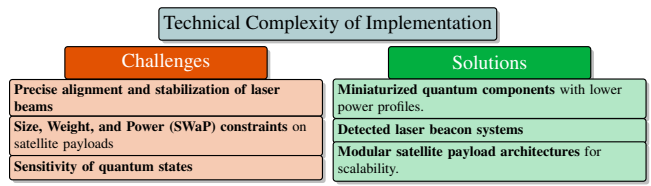

The technical complexity of the implementation represents a substantial challenge in the field of QSC as shown in Fig.~\ref{fig_challenge_7}. This complexity stems from the need to maintain \textbf{precise alignment and stabilization of laser beams} between rapidly moving satellites, in addition to the inherent fragility of quantum states, which are highly susceptible to environmental noise and disturbances \cite{madani2025quantum}. \textbf{The generation, detection, and manipulation of quantum states} require advanced, highly sensitive instrumentation, including single-photon detectors that must operate with high efficiency and low dark count rates \cite{jennewein2018towards}. These components must also be \textbf{miniaturized and ruggedized} to meet the physical and environmental constraints of satellite platforms for long-term operational reliability. A particularly critical aspect is \textbf{the maintenance of a stable optical link}, which depends on extremely accurate pointing and tracking systems. Even minor misalignments or jitter can result in photon loss, severely degrading communication quality and reliability \cite{madani2025quantum}. This is especially relevant when transmitting single photons or entangled photon pairs, where the margin for error is minimal.
To address these technical challenges, researchers are investigating a range of mitigation strategies. For instance, \textbf{dedicated laser beacon systems} are being employed to facilitate initial alignment and maintain link stability, while self-compensating receiver architectures are being developed to adaptively correct for alignment drift \cite{Orsucci2024AssessmentOP}. Insights and techniques from classical laser communication systems are also being adapted and optimized to support quantum applications. In parallel, solutions such as \textbf{advanced shielding, error correction techniques, and high-precision tracking mechanisms} are being explored for robustness under space conditions \cite{madani2025quantum}. Furthermore, the adoption of \textbf{hybrid space–terrestrial quantum communication networks}, which integrate satellites and ground stations equipped with optical telescopes alongside metropolitan-scale fiber-optic infrastructure. This offers a promising avenue for improving system performance and scalability \cite{Haldar2022TowardsGT}.

\subsection{High Costs}
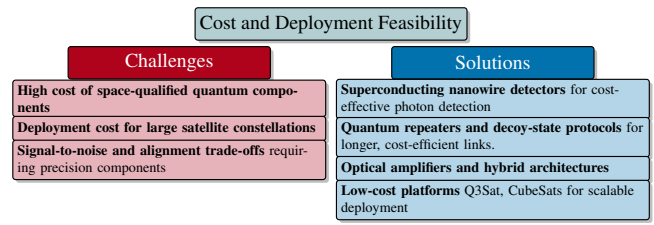
\begin{figure}[ht!]
    \definecolor{colorbraun}{HTML}{af021b}
\definecolor{colorlila}{HTML}{0272af}
\definecolor{colorgreen}{HTML}{005F69}
\tikzset{box/.style={rectangle, draw=black, rounded corners=2pt, thick},
    param/.style={box, font=\normalsize, minimum width=4cm, minimum height=0.6cm, text width=7.5cm, align=left, drop shadow},
    header/.style={box, font=\Large, text=black, minimum width=5cm, minimum height=0.8cm, drop shadow},
    icon/.style={circle, draw, thin, minimum size=6mm}}

\centering
\resizebox{1.0\columnwidth}{!}{
\begin{tikzpicture}[y=-0.8cm]

% Column headers
\node[header, fill=colorgreen!30] at (0,-0.2) {Cost and Deployment Feasibility};
\node[header, text=white, fill=colorlila] at (4cm,1) {Solutions};
\node[header, text=white, fill=colorbraun] at (-4cm,1) {Challenges};

% Challenges
\node[param, fill=colorbraun!30] at (-4,2.1) {\textbf{High cost of space-qualified quantum components}};
\node[param, fill=colorbraun!30] at (-4,3.03) {\textbf{Deployment cost for large satellite constellations}};
\node[param, fill=colorbraun!30] at (-4,4.0) {\textbf{Signal-to-noise and alignment trade-offs} requiring precision components};

% Solutions
\node[param, fill=colorlila!30] at (4,2.12) {\textbf{Superconducting nanowire detectors} for cost-effective photon detection};
\node[param, fill=colorlila!30] at (4,3.3) {\textbf{Quantum repeaters and decoy-state protocols} for longer, cost-efficient links.};
\node[param, fill=colorlila!30] at (4,4.27) {\textbf{Optical amplifiers and hybrid architectures}};
\node[param, fill=colorlila!30] at (4,5.24) {\textbf{Low-cost platforms} Q3Sat, CubeSats for scalable deployment};

\end{tikzpicture}
}
    \caption{Cost and deployment feasibility}
    \label{fig_challenge_8}
\end{figure}

The high costs of QSC systems pose a significant challenge to the widespread adoption and deployment of QKD technology. One of the main reasons for the high costs is the need for\textbf{ high-quality optical components}, such as single-photon detectors and laser sources, which are expensive and difficult to miniaturize \cite{Gozzard2021VulnerabilityOS}. Furthermore, the design of QKD systems for satellite applications requires careful consideration of factors such as signal-to-noise ratio (SNR), transmission loss, and beam diffraction, all of which affect the efficiency and performance of the system \cite{Zhang2023EndtoendDF}. Additionally, the use of free space channels for QKD can cause issues such as atmospheric noise and misalignment of the beam, which can further increase costs \cite{Tsai2021QuantumKD}. Fig.\ref{fig_challenge_8} illustrates an enumeration of the primary factors that contribute to the elevated costs associated with QSC systems, as well as the existing solutions that address these issues. 

However, researchers have proposed several solutions to mitigate the high costs of QSC. One approach is to use more efficient and compact optical components, such as \textbf{superconducting nanowire single-photon detectors}, which can reduce the costs of QKD systems \cite{Zhang2023EndtoendDF}. Another approach is to use \textbf{optical amplifiers} for improving SNR and increasing the communication distance, which can also reduce costs \cite{Tsai2021QuantumKD}. Moreover, the use of \textbf{quantum repeaters} can also help reduce the costs of long-distance QKD by allowing for the distribution of entangled photons over longer distances \cite{gundogan2024time}. Additionally, the development of new QKD protocols, such as the decoy state protocol, can also help to reduce costs by minimizing the number of photons required for the generation of secure keys \cite{Neumann2017Q3SatQC}. Furthermore, the use of \textbf{CubeSat platforms}, which are small and low-cost satellites, can also help to reduce the costs of QSC \cite{Zhang2023EndtoendDF}. The \textbf{Q3Sat mission}, for example, has demonstrated the feasibility of using a CubeSat platform for QKD, and has shown that it is possible to achieve secure key generation with a modest amount of resources \cite{Neumann2017Q3SatQC}.

\section{State-of-the-Art Research}
\label{Sec_stateOfArt}

\begin{figure*}[ht!]
\begin{center}
\includegraphics[width=1.0\textwidth]{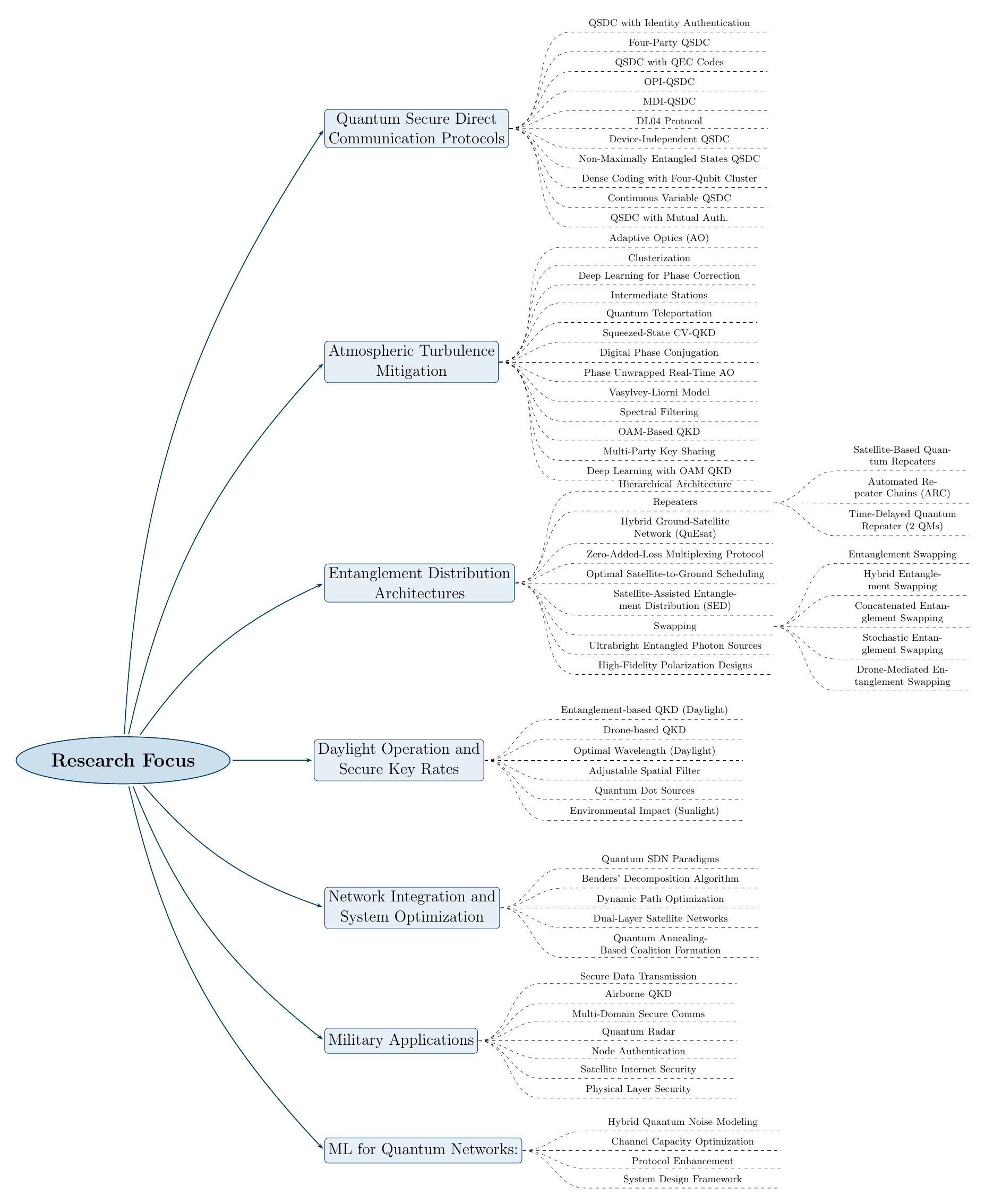}\\
\end{center}
\caption{Current research focus.}
\label{fig_currentfocus}
\end{figure*}

% \begin{figure*}[t]
%     \input{fig_research_focus_n}
%     \caption{Current research focus}
%     \label{fig_currentfocus}
% \end{figure*}

% \subsection{Current Research Focus}
More than 120 recent articles have converged on several key technical aspects and advances in QSC systems. Research analysis and abstract clustering reveal the following technical aspects that are frequently studied, as shown in Fig.\ref{fig_currentfocus}:

\subsection{Quantum Secure Direct Communication Protocols}
The development of quantum secure direct communication (QSDC) protocols is advancing rapidly, with protocols such as DL04 demonstrating significant improvements. QSDC is a quantum communication protocol that ensures information-theoretic security by using quantum states to transmit information directly. Unlike traditional QKD, QSDC does not require key exchange but instead encodes messages directly into quantum states. The evolution of QSDC has been driven by the need to address practical challenges such as noise robustness, long-distance communication, and security against eavesdropping attacks. Table~\ref{tab:qsdc_protocols} compares the current QSDC protocols. 

\begin{table*}[ht!]
    \centering
    \caption{Comparison of QSDC Protocols}
    \label{tab:qsdc_protocols}
    \begin{tabularx}{\textwidth}{XXX}
    % {|l|l|l|}
    \toprule
    \textbf{Protocol} & \textbf{Key Features} & \textbf{Advantages} \\ \midrule
    MDI-QSDC & Uses hyperentanglement and quantum covert channels & Eliminates measurement device vulnerabilities; high channel capacity \\        
        OPI-QSDC & Leverages single-photon interference & Doubles communication distance; no need for quantum memory \\        
        QSDC with QEC Codes & Incorporates quantum error correction codes & Reduces transmission errors; increases secret message capacity \\
        Four-Party QSDC & Utilizes hyperentangled Bell states in multiple degrees of freedom & Enables multi-party communication; high information transmission rates \\
        QSDC with Identity Authentication & Integrates GHZ states for identity verification & Resists common attacks; high transmission efficiency \\
        DL04 Protocol & Achieves high transmission rate (500 bps) with low QBER & Efficient and reliable for QSC\\
        Device-Independent QSDC & Does not rely on the trustworthiness of devices & Secure against eavesdropping; implementable with existing technologies \\
        Non-Maximally Entangled States QSDC & Uses non-maximally entangled states & Simple implementation; reduces leaked bits \\
        Dense Coding with Four-Qubit Cluster States & Employs dense coding of four-qubit cluster states & Bidirectional communication; novel quantum channel \\
        Continuous Variable QSDC (Gaussian States) & Uses Gaussian states for long-distance communication & Enables quantum dialogue and voting; secure over long distances \\
        Quantum Direct Communication with Mutual Authentication & Uses quantum Bell states and transformations for authentication & Ensures mutual authentication; authenticator has no access to secret message \\     
    \bottomrule
    \end{tabularx}
\end{table*}

Quantum Secure Direct Communication (QSDC) protocols have evolved rapidly, introducing innovative techniques to enhance security, efficiency, and practicality. One major breakthrough is the measurement-device-independent QSDC (MDI-QSDC) protocol, which removes the need to trust measurement devices by using entangled photon pairs and single photons \cite{zhou2020measurement, pan2025measurement}. Advances in hyperentanglement, leveraging multiple degrees of freedom such as polarization and orbital angular momentum, further improve channel capacity and security. Similarly, the one-photon-interference QSDC (OPI-QSDC) protocol addresses the limitations of quantum memory and ideal photon sources, doubling communication distance and making it highly suitable for satellite-based communication \cite{li2024one}.

To improve noise resilience and reliability, QSDC protocols increasingly integrate quantum error correction (QEC) techniques like repetition codes \cite{ding2025quantum}. Multi-party extensions, such as the four-party hyperentangled Bell-state protocol, allow independent, secure communication among multiple users with enhanced throughput \cite{guo2024four}. For identity authentication, protocols using GHZ states ensure only authorized parties decode the transmitted messages while maintaining high efficiency and resistance to attacks \cite{zhou2023quantum}. The DL04 protocol remains widely studied, achieving a 500 bps rate with low QBER in satellite QKD scenarios \cite{pan2023free}.

Device-independent QSDC approaches enhance security by removing trust in hardware, such as protocols using non-maximally entangled states to reduce information leakage \cite{Vijayaraj2019QuasiquantumSD}. Other innovations include bidirectional QSDC with four-qubit cluster states, enabling efficient dense coding between transmitter and receiver \cite{Chauhan2022BidirectionalQS}.

Applications have also broadened: continuous-variable QSDC with Gaussian states supports long-distance secure tasks like quantum dialogue and voting \cite{Srikara2019ContinuousVD}, while protocols incorporating mutual authentication via Bell states and unitary transformations ensure secure message transmission without exposing content \cite{Yen2009QuantumDC}.

\subsection{Atmospheric Turbulence Mitigation}

\begin{table*}[h]
    \centering
    \caption{Comparison of Turbulence Mitigation Strategies in QSC}
    \label{tab:turbulence_mitigation}
    \begin{tabularx}{\textwidth}{XXX}
        \toprule
        \textbf{Strategy} & \textbf{Mechanism} & \textbf{Key Benefits} \\ \midrule
        AO & Dynamically corrects wavefront distortions caused by atmospheric turbulence. & Improves coupling efficiency, reduces crosstalk, and enables real-time correction. \\ \hline
        Clusterization & Divides measurement data into clusters to suppress fading noise in CV-QKD. & Enhances robustness against large fading noise and finite-size effects. \\ \hline
        Deep Learning for Phase Correction & Uses CNNs to estimate phase corrections from intensity measurements. & Simplifies implementation and reduces reliance on complex phase-measuring equipment. \\ \hline
        Intermediate Stations & Acts as relay points to reduce turbulence effects in uplink and downlink scenarios. & Improves fidelity of entanglement distribution and quantum teleportation. \\ \hline
        Quantum Teleportation & Enables deterministic transfer of quantum states, reducing uplink limitations. & Overcomes turbulence and background radiation constraints in ground-to-satellite links. \\ \hline 
        Squeezed-State CV-QKD & Reduces excess noise and enhances robustness against channel attenuation. & Allows for lower system clock rates and smaller aperture sizes. \\ \hline
        Digital Phase Conjugation & Uses a probe beam to characterize turbulence and corrects distortions via phase modulation. & Effective in suppressing turbulence over long free-space links (e.g., 340 m). \\ \hline
        Phase Unwrapped Real-Time AO & Corrects turbulence-induced phase errors in OAM-based QKD using real-time AO. & Reduces errors in weak turbulence and establishes secure channels. \\ \hline
        Vasylvey-Liorni Model & Adapts beam profiles (spherical/elliptic) based on humidity levels. & Reduces QBER by up to 42\% in LEO satellite links. \\ \hline
        Spectral Filtering & Filters out solar noise during daylight operations. & Enhances system performance under high background radiation. \\ \hline
        OAM-Based QKD & Uses OAM of light for high-dimensional encoding. & Increases data rates by utilizing higher-dimensional Hilbert space. \\ \hline
        Multi-Party Key Sharing & Hybrid entangled states of single photons in high dimensions. & Enables secure key distribution among multiple parties. \\ \hline
        Deep Learning with OAM QKD & Integrates deep learning for phase correction in OAM-encoded QKD. & Enhances robustness against turbulence, even in high wind velocity regimes. \\ 
        \bottomrule
    \end{tabularx}
\end{table*}

Recent research on atmospheric turbulence mitigation in Quantum Satellite Communication (QSC) has yielded advanced techniques to improve the reliability and security of quantum links \cite{pugh2020adaptive}. Table~\ref{tab:turbulence_mitigation} provides a comparative overview of these strategies. A key solution is adaptive optics (AO), which dynamically corrects wavefront distortions using components such as wavefront sensors, deformable mirrors, and control units \cite{Koudia2024PhysicalLA}. AO enhances signal-to-noise ratio (SNR) and reduces quantum bit error rate (QBER), particularly in free-space QKD, by stabilizing shaped beams like vector Bessel-Gauss beams, which mitigate scintillation and beam wander. In satellite-ground links, AO has improved fiber coupling efficiency, boosting secret key rates by several hundred bits per second \cite{acosta2024increasing}, with further gains up to 7 dB when combined with laser guide stars \cite{pugh2020adaptive}.

In continuous-variable QKD (CV-QKD), turbulence-induced fading noise degrades security. Clustering methods based on probability distributions of transmittance (PDT) mitigate this issue, reducing noise while managing finite-size effects \cite{derkach2020applicability, zuo2023suppression}. Similarly, deep learning approaches, particularly convolutional neural networks (CNNs), have enabled phase correction without classical reference pulses, achieving secure key rates in satellite-ground CV-QKD links \cite{long2024phase}. When integrated with spatial light modulators, these models have also enhanced orbital angular momentum (OAM)-encoded QKD, improving mode purity and extending secure communication distances, even in turbulent or high-wind conditions \cite{wang2021integrating}.

Other techniques include digital phase conjugation, which uses probe beams and phase-only spatial light modulators to suppress turbulence effects, demonstrating success over a 340 m free-space link \cite{Zhao2020PerformanceOR}. Additionally, phase-unwrapped AO correction has proven effective in reducing turbulence-induced errors in OAM-QKD channels \cite{Tao2021MitigatingTE}. Hybrid models such as the Vasylyev-Liorni model adapt beam profiles dynamically, reducing QBER by up to 42\% under varying atmospheric conditions \cite{ntanos2021leo}. Spectral filtering systems have also been critical in reducing solar noise during daylight operations.

High-dimensional encoding using OAM modes offers substantial potential for increasing QKD data rates \cite{Wang2018TowardsPH, Wang2019DetectingOA, Oliveira2019ExperimentalQS, Wang2020SatelliteToEarthQK}. Although OAM states are sensitive to turbulence, AO systems with real-time correction have significantly reduced spatial-mode crosstalk \cite{scarfe2025fast}. Experimental demonstrations have shown secure OAM-encoded entanglement distribution over a 3 km free-space optical link \cite{Oliveira2019ExperimentalQS}, achieving high key generation rates and improved error resilience. Integration of deep learning-based phase correction with OAM-QKD further enhances robustness in challenging atmospheric conditions \cite{wang2021integrating}.

Moreover, quantum teleportation and entanglement distribution have benefited from intermediate relay stations for beam refocusing, improving fidelity in both uplink and downlink paths \cite{gonzalez2024satellite}. In CV teleportation, optimized entangled resources have maintained secure transmission even under strong turbulence \cite{zuo2021overcoming}. Finally, the use of squeezed states in CV-QKD improves robustness against channel attenuation and noise, allowing lower clock rates and smaller apertures compared to coherent-state protocols \cite{derkach2020applicability}. Collectively, these innovations highlight a multidisciplinary approach—combining AO, deep learning, high-dimensional encoding, and hybrid models—to overcome atmospheric turbulence, paving the way for more reliable, scalable, and high-capacity QSC systems.

\subsection{Entanglement Distribution Architectures}
\begin{table*}[h]
    \centering
    \caption{Entanglement Distribution Techniques in QSC}
    \label{tab:entanglement_distribution}
    \begin{tabularx}{\textwidth}{XXX}
        \toprule
        \textbf{Technique} & \textbf{Description} & \textbf{Key Feature} \\ \midrule
        Hierarchical Architecture & Organizes nodes hierarchically to optimize entanglement distribution efficiency. & Improves efficiency by 11.5\% and routing performance by 37.3\%. \\ \hline
        Hybrid Ground-Satellite Network (QuEsat) & Combines ground fiber networks with LEO satellites using adjustable lens arrays. & Enables dynamic photon lightpath construction with high long-distance efficiency. \\ \hline
        Satellite-Based Quantum Repeaters & Uses trapped atomic qubits as photon sources and quantum memories. & Achieves high-fidelity entanglement over inter-continental distances without cryogenics. \\ \hline
        Entanglement Swapping & Distributes entanglement without physical transport of particles. & Effective in satellite systems; avoids exponential fiber loss. \\ \hline
        Hybrid Entanglement Swapping & Combines DV and CV systems for flexible teleportation. & Retains higher entanglement quality than direct DV distribution. \\ \hline
        Concatenated Entanglement Swapping & Uses multiple swapping steps for extreme distances. & Enables global-scale entanglement but requires parameter optimization. \\ \hline
        Zero-Added-Loss Multiplexing Protocol & Distributes entanglement via separable states in noisy environments. & Robust against noise; suitable for global-scale networks. \\ \hline
        Optimal Satellite-to-Ground Scheduling & Optimizes transmission policies for maximum entanglement rates. & Balances resource constraints (e.g. satellite/ground station limitations). \\ \hline
        Stochastic Entanglement Swapping Analysis & Models probabilistic swapping in quantum repeaters. & Provides cost-effective expressions for system performance. \\ \hline
        SED & Reduces qubit consumption via optimized satellite links. & Cuts qubit use by 38\% compared to traditional methods. \\ \hline
        ARC & Integrates automated repeaters for scalable networks. & Extends coverage and improves reliability. \\ \hline
        Drone-Mediated Entanglement Swapping & Uses drones as mobile repeater networks. & Enhances network flexibility and coverage in remote areas. \\ \hline
        Time-Delayed Quantum Repeater (2 QMs) & Employs two quantum memories on a satellite for stored qubit swapping. & Achieves 3x higher key rates with reduced memory requirements. \\ \hline
        Ultrabright Entangled Photon Sources & Optimized polarization/frequency hyper-entanglement for free-space transmission. & Overcomes detector bandwidth limits; compatible with Si-SPADs. \\ \hline
        High-Fidelity Polarization Designs & Compensates for satellite motion-induced polarization shifts. & Achieves extinction ratio of 887:1 and fidelity > 0.995. \\ \bottomrule
    \end{tabularx}
\end{table*}

Recent advancements in entanglement distribution architectures in Quantum Satellite Communication (QSC) have focused on scalable and robust solutions for global quantum networks \cite{He2024BuildingAH}. Table~\ref{tab:entanglement_distribution} compares these techniques. A key innovation is the hierarchical architecture, improving entanglement distribution efficiency by 11.5\% and routing performance by 37.3\% compared to traditional models \cite{He2024BuildingAH}. Similarly, the QuESat hybrid architecture integrates ground-based fiber networks with low-Earth-orbit (LEO) satellites, enabling near-vacuum photon lightpaths for highly efficient long-distance transmission \cite{Gu2025QuESatSQ}.

Quantum repeaters are central to extending distribution range. Satellite-based repeaters using trapped atomic qubits function as photon sources and quantum memories, providing high-fidelity entanglement over inter-continental distances without cryogenic systems \cite{tubio2024satellite, Tubio2024SatelliteassistedQC}. Combining quantum memories with satellite links supports global-scale quantum key distribution (QKD) and enhances entanglement rates for line-of-sight connections \cite{liorni2021quantum, boone2015entanglement}. Protocols such as MDI-QKD \cite{Lewis2021QuantumCI} and memory-assisted QKD \cite{Gndoan2020ProposalFS} further boost key rates and enhance device-independent security. Despite challenges like channel noise modeling and optical loss, quantum memories in satellites provide a pathway toward hybrid satellite–fiber networks interconnecting metropolitan areas \cite{Fittipaldi2024EntanglementSI}.

Entanglement swapping enables distribution without physically transporting particles, significantly reducing fiber loss. This technique, demonstrated in both uplink and downlink links \cite{floyd2023long}, supports hybrid DV–CV protocols for teleporting arbitrary states with higher entanglement quality under lossy conditions \cite{do2019hybrid}. For ultra-long distances, concatenated swapping extends range but requires careful optimization to balance pair production rates and detector efficiency \cite{khalique2017long}.

Other methods include entanglement distribution through separable states using photon-multiplexing-inspired protocols, which offer resilience against noise \cite{Campbell2024EntanglementDT}. Optimization studies have explored scheduling policies for maximizing aggregate distribution rates while considering satellite and ground station resource constraints \cite{Panigrahy2022OptimalED}. Stochastic analyses of quantum repeaters have also simplified performance modeling with Markov chains, yielding analytical performance expressions \cite{Vardoyan2019OnTS}.

Emerging strategies include satellite-assisted entanglement distribution (SED), reducing qubit consumption by 38\% \cite{sondergaard2025satellite}, and automated repeater chains (ARC) or drone-mediated swapping for network extension. A novel dual-quantum-memory satellite design acts as a time-delayed repeater node, physically transporting stored qubits to improve secure key generation by up to three-fold \cite{gundoǧan2024time}.

High-performance entangled photon sources are critical for long-distance links. Ultrabright polarization-entangled photon pair sources optimized for space transmission leverage hyper-entanglement to overcome detector bandwidth limits and match wavelengths compatible with space-ready Si-SPADs \cite{brambila2023ultrabright}. Advanced polarization designs for ground-to-satellite links achieve extinction ratios of 887:1, while compensation schemes ensure entanglement fidelity exceeding 0.995, even with satellite motion \cite{han2020polarization}.

%Collectively, these innovations — hierarchical and hybrid architectures, quantum repeaters with memories, advanced swapping protocols, and ultrabright photon sources — pave the way for a global quantum Internet. These systems promise high scalability, improved key rates, and robust performance under challenging conditions, positioning satellite-based entanglement distribution as a cornerstone for future quantum communication, navigation, and sensing infrastructures.

\subsection{Daylight Operation and Secure Key Rates}
Secure key generation rates have been shown to reach 5.7 kbps under daytime conditions using QKD systems based on entanglement, as noted by \cite{krvzivc2023towards}. Furthermore, experimental advances in drone-based QKD have achieved secret key rates of 8.48 kHz over 200 m links according to \cite{tian2024experimental}. 
\begin{table}[h]
    \centering
    \caption{Secure Key Generation Rates in QSC}
    \label{tab:secure_key_aspects}
    \begin{tabularx}{\columnwidth}{p{1.8cm}p{2.8cm}p{2.9cm}}
    % {|p{3.9cm}|p{5.5cm}|p{5.5cm}|}
    \toprule
    \textbf{Key Aspect} & \textbf{Findings/Technique} & \textbf{Performance} \\ \midrule
    Entanglement-Based QKD (Daylight) & QKD systems using entanglement for secure key generation. & Achieves 5.7 kbps in daytime conditions. \\ \hline
    Drone-Based QKD & Experimental drone-based QKD over free-space links. & 8.48 kHz secret key rate over 200 m. \\ \hline
    Optimal Wavelength (Daylight) & H\textsubscript{$\alpha$} Fraunhofer line at 656 nm for satellite QKD. & 10× higher secret key rate vs. C-band; lower QBER in daylight. \\ \hline
    Adjustable Spatial Filter & Active optimization of field of view to match link conditions. & Improves secure key rates in daylight and nighttime. \\ \hline
    Quantum Dot Sources & Entanglement-based QKD using quantum dots in daylight. & 106 bit/s sifted key rate, 7.16\% QBER. \\ \hline
    Environmental Impact (Sunlight) & Study of sunlight effects on QKD performance. & Sifted key rate unchanged, but QBER increases slightly with sunlight intensity. \\ \bottomrule
    \end{tabularx}
\end{table}

Table~\ref{tab:secure_key_aspects} shows the current key aspects in the secure key generation rates in QSC systems. One of the key findings is that the ideal wavelength for satellite-based QKD during daylight conditions is the H$\alpha$ Fraunhofer line at 656 nm, which provides a secret key rate that is more than 10 times higher than any source operating at C-band \cite{Abasifard2023TheIW}. This is because the H$\alpha$ Fraunhofer line has a lower QBER during daylight, which allows for a higher secret key rate. Another important finding is that active optimization of the system field of view with the ever-changing link conditions can improve the total secure key rate output \cite{Kri2023AdjustableSF}. This can be achieved using an adjustable spatial filter, which can substantially improve the system's performance in daylight and night-time conditions. In addition, the use of quantum dot sources has been shown to be effective in entanglement-based QKD in daylight conditions \cite{BassoBasset2022DaylightEQ}. For example, a recent experiment demonstrated the feasibility of entanglement-based QKD using a quantum dot source in daylight, with a sifted key rate of 106 bit/s and a QBER of 7.16(2)\%. 

The impact of environmental conditions on the performance of QKD systems in daylight has also been studied. For example, one experiment found that the sifted key rate does not show any significant correlation with the level of sunlight, but the QBER (CHSH parameter) does slightly increase (decrease) with higher level of sunlight. This is consistent with the level of background detection events due to incomplete filtering of external light \cite{BassoBasset2022DaylightEQ}.

%In general, recent work on daylight operation and secure key rates in QSC has highlighted the importance of optimizing the field of view of the system and using appropriate wavelengths and sources to achieve high secret key rates under daylight conditions.

\subsection{Network Integration and System Optimization}
Recently, significant work has been done on network integration and system optimization in QSC, including the co-integration of satellite and terrestrial networks using quantum software-defined networking paradigms and dynamic path optimization algorithms to ensure robust performance under environmental variability \cite{Aguado2019TheEO, AlHraishawi2022CharacterizingAU, bakker2024best}. Table~\ref{tab:network_integration} categorizes the current technologies for network integration and system optimization in QSC systems.

\begin{table}[h]
    \centering
    \caption{Network Integration \& System Optimization in QSC}
    \label{tab:network_integration}
    \begin{tabularx}{\columnwidth}{p{1.8cm}p{2.8cm}p{2.9cm}}
    \toprule
    \textbf{Category} & \textbf{Approach/Technology} & \textbf{Benefits/Performance} \\
    \midrule
    Quantum SDN Paradigms & Software-defined networking for flexible quantum network configuration & Enables dynamic management of time-varying satellite networks with thousands of nodes \\
    \hline
    Benders’ Decomposition Algorithm & MILP problem decomposition for efficient routing & Solves complex resource allocation problems in large-scale quantum networks \\
    \hline
    Dynamic Path Optimization & Adaptive routing algorithms for changing environmental conditions & Optimizes quantum signal transmission in variable link conditions \\
    \hline
    Dual-Layer Satellite Networks & Integration of LEO/GEO satellites with MEC and caching & Enhances service efficiency through coordinated resource allocation \\
    \hline
    Quantum Annealing-Based Coalition Formation & Optimal partitioning of LEO satellites into communication coalitions & Solves NP-Hard optimization problems for reliable satellite groupings \\
    \bottomrule
    \end{tabularx}
\end{table}

Efficient integration of satellite and terrestrial networks remains a key challenge in QSC, requiring advanced routing and resource allocation algorithms \cite{Li2025ResourceAA}. Software-defined networking (SDN) offers flexible, programmable configurations that enhance performance and monitoring, making it ideal for dynamic quantum topologies with thousands of nodes. Techniques like Benders' decomposition (BD)-based algorithms efficiently break down complex MILP problems \cite{Zhang2024EfficientER}, while dynamic path optimization adapts to environmental changes, improving quantum signal routing. Recent strategies integrate MEC, network caching, and SDN in dual-layer LEO-GEO networks to enhance service efficiency \cite{Li2025ResourceAA}. Beyond SDN, quantum annealing algorithms have been proposed for coalition formation among LEO satellites \cite{Venkatesh2024QuantumAA}, tackling the NP-Hard problem of optimal satellite partitioning to ensure reliable communications. These advancements collectively address resource allocation, interference coordination, and robustness challenges, laying the groundwork for next-generation quantum communication infrastructures to support 6G and beyond.

\subsection{Military Applications}
In recent years, there has been significant progress in the development of QSC for military applications. Table~\ref{tab:quantum_satellite_military} shows the current technologies and approaches in military applications for QSC systems. 
\begin{table}[h]
    \centering
    \caption{QSC for Military Applications}
    \label{tab:quantum_satellite_military}
    \begin{tabularx}{\columnwidth}{p{1.8cm}p{2.8cm}p{2.9cm}}
    \toprule
    \textbf{Application} & \textbf{Technology/Approach} & \textbf{Key Advantages} \\ \midrule
    Secure Data Transmission & QKD systems & Provides unconditional security through quantum principles \\ \hline
    Airborne QKD & Ground-to-aircraft QKD demonstrations & Validates QKD for mobile military platforms \\ \hline
    Multi-Domain Secure Comms & Quantum links between ships, subs, satellites \& ground stations & Enables secure cross-domain communication networks \\ \hline
    Quantum Radar & Space-based quantum radar systems & Offers superior detection and tracking sensitivity vs classical radar \\ \hline
    Node Authentication & Quantum entanglement-based authentication & Uses quantum measurement noise for secure node verification \\ \hline
    Satellite Internet Security & Quantum encryption protocols & Provides anti-eavesdropping protection for space-based networks \\ \hline
    Physical Layer Security & Precoding, cooperative jamming, and PLA techniques & Enhances security at the physical transmission layer \\ \bottomrule
    \end{tabularx}
\end{table}

Ensuring secure data transmission in satellite communication is a major challenge due to the risk of signal interception. Quantum Key Distribution (QKD) leverages quantum mechanics to provide unconditional security \cite{Krelina2021QuantumTF}. QKD has been validated in various experiments, including a ground-to-airborne receiver demonstration \cite{Pugh2016AirborneDO}, highlighting its potential for satellite-based communication. Research is also exploring quantum communication for military applications, enabling secure links between ships, satellites, underwater vessels, and ground stations \cite{Neumann2020QuantumCF}. Quantum radar is being investigated for space applications, offering superior detection and tracking sensitivity compared to classical radar. Additionally, quantum entanglement and measurement noise are proposed for satellite node authentication \cite{madani2025quantum}. In satellite internet, quantum encryption and PLS techniques—such as precoding, cooperative jamming, relay selection, and PLA—are emerging as promising solutions for secure, reliable, and anti-eavesdropping communications \cite{Zhang2023ASO}, paving the way for next-generation secure satellite networks.

\subsection{Machine Learning (ML) for Quantum Networks} 
The recent work on exploring machine learning to optimize satellite-based quantum communication in QSC has focused on several key areas, including modeling hybrid quantum noise, optimizing quantum communication systems, and developing novel machine learning algorithms for satellite-based quantum communication. Table~\ref{tab:ML_opti} shows the list of focus areas in using machine learning (ML) for the optimization of the QSC system.
\begin{table}[h]
    \centering
    \caption{Machine Learning for QSC Optimization}
    \label{tab:ML_opti}
    \begin{tabularx}{\columnwidth}{p{1.4cm}p{1.5cm}p{1.8cm}p{2.1cm}}
    \toprule
    \textbf{Focus Area} & \textbf{ML Technique} & \textbf{Key Innovation} & \textbf{Impact} \\ \midrule
    Hybrid Quantum Noise Modeling & Quantum Poissonian + AWGN noise synthesis & Unified model for realistic channel conditions & 38\% more accurate capacity estimation vs. Gaussian-only models \\ \hline
    Channel Capacity Optimization & GMM + EM clustering & Dynamic noise clustering for QKD systems & 25\% higher SKR in LEO-to-ground tests \\ \hline
    Protocol Enhancement & Phase-matching MDI-QKD analysis & Asymptotic key rate evaluation under satellite channel conditions & Validates feasibility for 500-1200 km orbits with $>$1 kbps secure key rates \\ \hline
    System Design Framework & SNR-SKR correlation mapping & ML-generated performance landscapes for satellite QKD & Enables adaptive parameter tuning for varying atmospheric conditions \\ \bottomrule
    \end{tabularx}
\end{table}

The work in \cite{Chakraborty2024AHN} introduces a hybrid quantum noise model integrating quantum Poissonian noise with classical AWGN, offering a more accurate representation of quantum channel capacity. Leveraging ML-generated clusters, the model proposed in \cite{Chakraborty2024AnUM} optimizes Gaussian quantum channels and reduces cluster complexity, enhancing visualization and capacity estimation. Novel ML algorithms, including Gaussian mixture models (GMM) and expectation maximization (EM), effectively model complex noise in satellite-based quantum communication, improving QKD performance. The hybrid approach also explores SKR variations with SNR, providing practical insights for system design. Additionally, the feasibility of phase-matching measurement-device-independent QKD (PM-MDI QKD) protocols for satellite channels is evaluated \cite{Dutta2025SatellitebasedCF}, demonstrating reliable key rates under typical noisy and loss-only conditions. This study advances both theoretical modeling and practical frameworks for robust, high-performance satellite-based quantum communication systems.

\section{Future Directions}
\label{Sec_future}
\begin{figure*}[ht!]
    \definecolor{leo-blue}{RGB}{64,115,158}
\definecolor{geo-orange}{RGB}{218,124,48}
\begin{tikzpicture}
% Define the block above
\node (top) [draw, fill=leo-blue, text width=12em, font=\sffamily\bfseries\footnotesize\color{white}, align=center, rounded corners=2pt] at (0, 0) {QSC Future Directions};

% Define the three forest diagrams
\node (tree1) at (-6.0, -3.625) {
\begin{forest}
basic/.style = {draw, thin, drop shadow},
my root/.style = {basic, rounded corners=0pt,fill=white, text width=12em,,align=left,font=\sffamily\scriptsize\sffamily, rounded corners=2pt},
upper style/.style = {basic, rounded corners=0pt, fill=leo-blue!10, text width=11.8em,,align=left,font=\sffamily\scriptsize, rounded corners=2pt},
lower style/.style = {basic, fill=leo-blue!90,rounded corners=0pt, text width=13em,font= \sffamily\bfseries\scriptsize\color{white}, rounded corners=2pt},
 for tree={grow'=0,folder,draw,
 where level=0{lower style}{},
 where level=1{upper style}{},
 where level=2{my root}{}
 },
[Global Quantum Internet
    [Satellite Constellations]
    [Hybrid Networks]
    [Integration with Classical Networks]
    [Quantum Repeaters]
]
\end{forest}
};

\node (tree2) at (0, -4) {
\begin{forest}
basic/.style = {draw, thin, drop shadow, rounded corners=2pt},
my root/.style = {basic, rounded corners=0pt,fill=white, text width=13em,,align=left,font=\sffamily\scriptsize\sffamily},
upper style/.style = {basic, rounded corners=0pt, fill=leo-blue!10, text width=14.7em,,align=left,font=\sffamily\scriptsize, rounded corners=2pt},
lower style/.style = {basic, fill=leo-blue!90,rounded corners=0pt, text width=15.8em,font= \sffamily\bfseries\scriptsize\color{white}, rounded corners=2pt},
 for tree={grow'=0,folder,draw,
 where level=0{lower style}{},
 where level=1{upper style}{},
 where level=2{my root}{}
 },
[Quantum Sensing and Metrology
    [Precision Measurements]
    [Enhanced Communication]
    [Integration with Communication Payloads]
    [Overcoming Technical Challenges]
    [Scalability and Network Formation]
]
\end{forest}
};

\node (tree3) at (6.0, -3.2) {
\begin{forest}
basic/.style = {draw, thin, drop shadow, rounded corners=2pt},
my root/.style = {basic, rounded corners=0pt,fill=white, text width=11em,,align=left,font=\sffamily\scriptsize\sffamily, rounded corners=2pt},
upper style/.style = {basic, rounded corners=0pt, fill=leo-blue!10, text width=13em,,align=left,font=\sffamily\scriptsize, rounded corners=2pt},
lower style/.style = {basic, fill=leo-blue!90,rounded corners=0pt, text width=14.2em,font= \sffamily\bfseries\scriptsize\color{white}, rounded corners=2pt},
 for tree={grow'=0,folder,draw,
 where level=0{lower style}{},
 where level=1{upper style}{},
 where level=2{my root}{}
 },
[Quantum Computing in Space
    [Space-Qualified Quantum Processors]
    [High-Dimensional Multipartite QC]
    [dvanced Payloads and System Design]
]
\end{forest}
};

% Draw arrows from the top block to each "Key Drivers" block
\draw[->, thick] (top.south) to (tree1.north);
\draw[->, thick] (top.south) -- (tree2.north);
\draw[->, thick] (top.south) to (tree3.north);

\end{tikzpicture}
    \caption{QSC future directions}
    \label{fig_future}
\end{figure*}
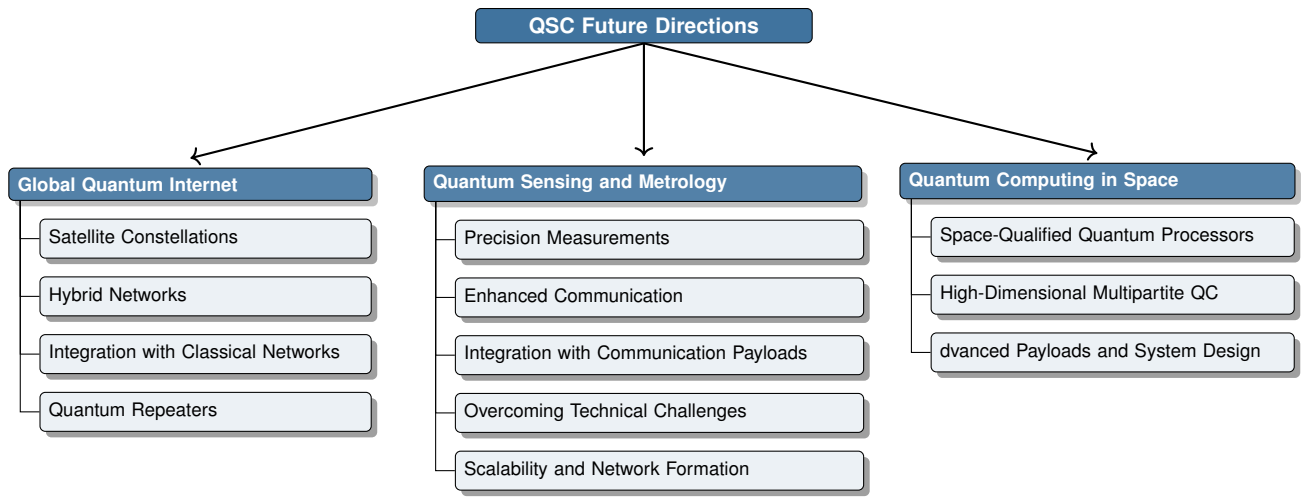

Based on the deep analysis of the current research focus, Fig.~\ref{fig_future} presents a visual overview of the anticipated future directions in QSC systems. 

\subsection{Global Quantum Internet}

The focus is on creating a network of quantum satellites interconnected with terrestrial quantum networks to enable secure global communication. The vision is to establish a robust and secure global communication infrastructure that takes advantage of the principles of quantum mechanics to ensure unparalleled security and efficiency. The research focus is on scalable architectures, integration with classical networks, and quantum repeaters, which are essential for realizing this vision. Here are the key aspects:

% Scalable Architectures
\subsubsection{Satellite Constellations} Researchers are exploring the optimal configurations of satellite constellations for improving QKD between ground stations. For example, the use of quantum relay satellites in Molniya orbits has been proposed to enhance communication efficiency and coverage. These high-eccentricity orbits allow satellites to extend their operational presence over targeted hemispheres, maximizing the reach of the quantum network.
\subsubsection{Hybrid Networks} This focus on developing hybrid terrestrial-satellite solutions to ensure seamless, ultra-secure connectivity. This involves integrating satellite-based quantum communication with existing terrestrial networks to take advantage of the benefits of both systems.
\subsubsection{Integration with Classical Networks} Future advancements will focus on seamless integration with classical communication networks. This will involve developing technologies that can bridge the gap between quantum and classical systems, allowing for the gradual transition to a fully quantum-secure communication infrastructure.
\subsubsection{Quantum Repeaters} To overcome distance limitations, quantum repeaters are essential for extending the range of quantum communication. They help recreate entanglement links over shorter segments, addressing the challenge of preserving entanglement across substantial distances. This is particularly important for satellite-based quantum communication, where the distance between satellites and ground stations can be significant. Innovative approaches are essential to implement quantum repeaters, including the use of entanglement swapping.

\subsection{Quantum Sensing and Metrology}

It is poised to revolutionize various fields by enhancing precision measurements and secure communication. Here are the key aspects:

\subsubsection{Precision Measurements} 
Satellites equipped with quantum sensors can achieve unprecedented accuracy in measuring gravity, time, and magnetic fields. For instance, quantum gravimeters and magnetometers can detect minute variations in gravitational and magnetic fields, enabling applications such as mineral exploration, climate monitoring, and geophysical surveys. Quantum atomic clocks can provide extremely precise time synchronization, which is crucial for global positioning systems (GPS) and other time-sensitive applications.

\subsubsection{Enhanced Communication} 
Quantum sensors integrated with communication payloads can improve the security and efficiency of satellite communication. Quantum entanglement swapping allows for secure and efficient transmission of information over long distances. This technology can also enable more precise measurements from sensors that can be fed directly into quantum computers for advanced data processing.

\subsubsection{Integration with Communication Payloads} 
The integration of quantum sensors with communication payloads is a key research focus. This involves developing compact, lightweight, and space-hardened quantum sensors that can be seamlessly integrated with satellite communication systems. For example, the QYRO project aims to create laser-based quantum gyroscopes for precise attitude control of satellites, which is crucial for maintaining high-quality communication links.

\subsubsection{Overcoming Technical Challenges} 
Researchers are working on mitigating noise sources and systematic effects that can degrade the performance of quantum sensors in real-world applications. This includes developing advanced error correction techniques, improving the stability and bandwidth of sensors, and exploring novel sensing modalities such as hybrid sensors and large-momentum transfer interferometry.

\subsubsection{Scalability and Network Formation} 
Another important direction is scaling up quantum sensing and communication systems to form large-scale networks. This involves developing protocols for quantum entanglement distribution and swapping over long distances, as well as creating robust and scalable quantum communication infrastructures.

\subsubsection{Role of Fisher Information in Quantum Metrology} 
At the core of quantum metrology lies the concept of Fisher information, which quantifies how much information an observable carries about an unknown parameter to be estimated. In quantum systems, the Quantum Fisher Information (QFI) serves as the ultimate bound on estimation precision, defining the so-called quantum Cramér–Rao bound. Leveraging entanglement and squeezed states allows quantum sensors to achieve sensitivities surpassing the classical shot-noise limit, approaching the Heisenberg limit. This makes QFI a fundamental tool for designing and optimizing satellite-based quantum sensors, enabling ultra-precise tasks such as gravitational wave detection, Earth observation, and timing synchronization in quantum satellite communication networks.

\subsection{Quantum Computing in Space}
Deploying quantum computers on satellites enables distributed quantum computing, where multiple quantum processing units (QPUs) are networked together. This setup supports the exchange of both quantum and classical information, enhancing computational capabilities. The applications of quantum computing in space are vast, including precise time synchronization, enhanced sensor precision, and secure quantum-enhanced applications like fault-tolerant systems and blind quantum computing. Here are the key aspects:

\subsubsection{Space-Qualified Quantum Processors} The development of space-qualified quantum processors is crucial to realizing these applications. These processors need to be robust enough to withstand the harsh conditions of space while maintaining high performance. Boeing's collaboration with HRL Laboratories on the Q4S mission is an example of efforts to develop such space-hardened technology.

\subsubsection{High-Dimensional Multipartite Quantum Communications} Future research will focus on the integration of high-dimensional multipartite quantum communications with sensing, computing, and intelligence for multiple users. This involves developing advanced quantum communication protocols and systems that can handle complex, multi-user scenarios.
\subsubsection{Advanced Payloads and System Design} There is a need for advanced quantum payloads, such as high-rate QKD systems, low-loss space-to-ground links, and intersatellite links. Developing compact and efficient receivers for ground stations is also essential to make quantum communication more accessible and practical.

\section{Conclusion}
\label{Sec_conc}

QSC is a promising technology to redefine secure communication on a global scale. This review has illuminated the remarkable progress in QSC, from early experiments such as Micius to sophisticated hybrid networks that integrate satellites and terrestrial systems. Key insights emerge: entanglement-based protocols offer unparalleled security, adaptive optics mitigate atmospheric turbulence, and AI-driven optimization unlocks new efficiencies. However, several challenges remain, particularly those related to atmospheric interference, signal stability, cost, and system scalability, which require continued innovation. The transformative potential of QSC is undeniable. It could underpin a quantum Internet, enable ultra-secure military and financial communications, and even support distributed quantum computing in space. As advances in miniaturization and error correction accelerate, the vision of a globally connected quantum network grows closer. For researchers, the message is clear: continued cross-disciplinary research and international collaboration are essential to overcome technical barriers and ensure the reliable deployment of this transformative technology.

\bibliographystyle{IEEEtran}
\bibliography{references}

\begin{IEEEbiography}
[{\includegraphics[width=1in,height=1.25in,clip,keepaspectratio]{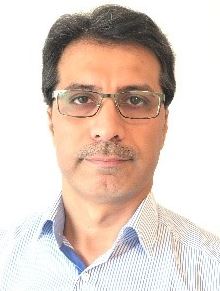}}]{Dr. Omar Alnaseri} (Senior Member, IEEE) is currently an adjunct lecturer at the DHBW Cooperative State University in Germany. Dr. Alnaseri holds B.Sc. and M.Sc. degrees in Electronics and Communications Engineering from the University of Technology in Baghdad, Iraq (awarded in 2000 and 2003, respectively). In 2015, he further distinguished himself by completing a Ph.D. in Optical Fiber Communications Engineering from the University of Paderborn, Germany. He built a high-speed end-to-end optical fiber transmission in the optical lab. at University of Paderborn. He has a special interest in deep learning and machine learning in (Quantum) communication systems. Dr. Alnaseri brings over 12 years of experience in the mobile/satellite communication sector, and 8 years of experience in embedded systems sector. This experience encompasses both research and industry roles, including research at the Technical University of Ilmenau, Germany, focusing on reinforcement learning algorithms for communication systems.
\end{IEEEbiography}
\vskip 0pt plus -1fil

\begin{IEEEbiography}[{\includegraphics[width=1in,height=1.25in,clip,keepaspectratio]{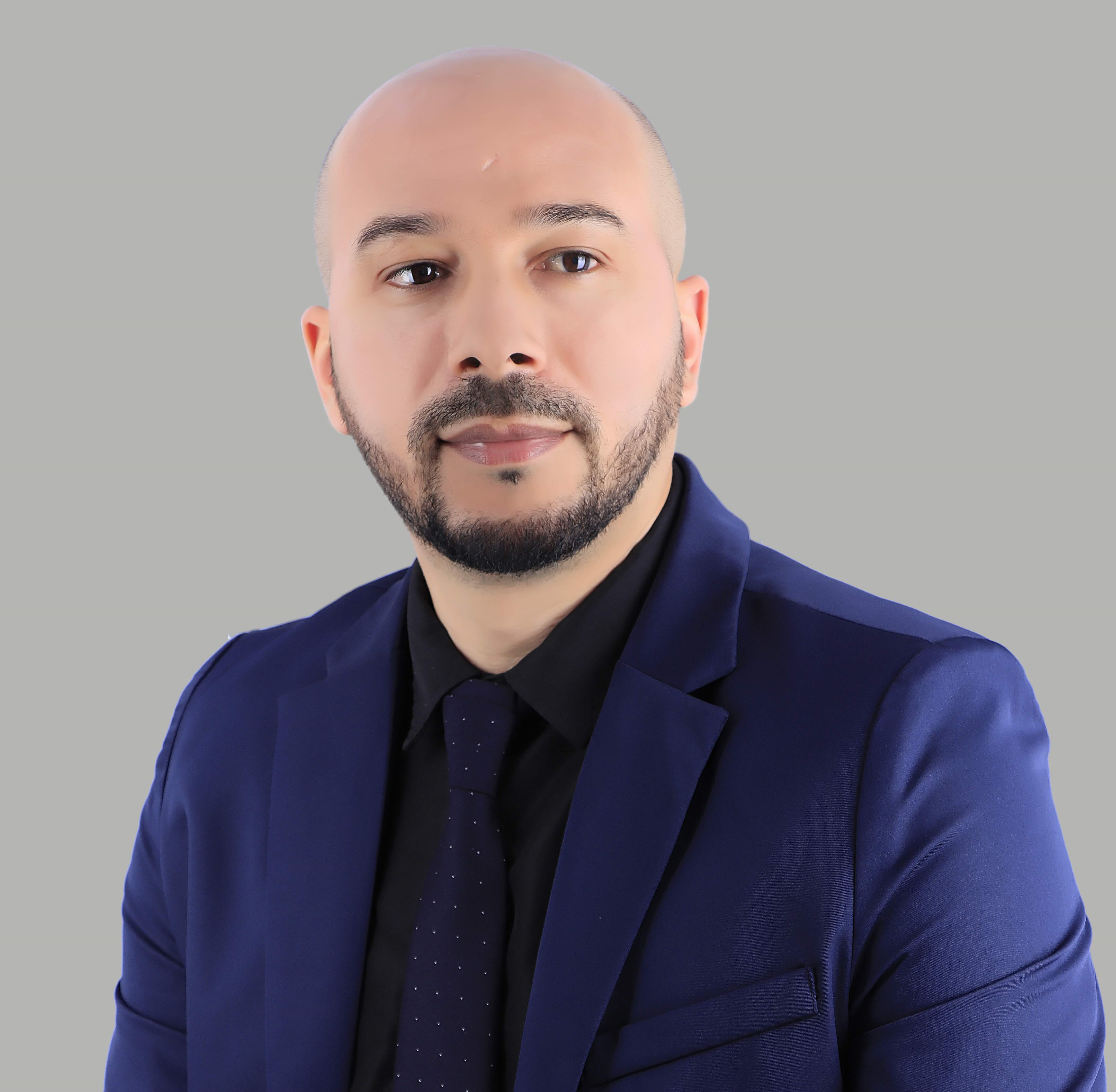}}]{Dr. Yassine Himeur} (\textit{Senior Member, IEEE}) is globally recognized for his significant contributions to Artificial Intelligence, especially in areas where AI intersects with cybersecurity, big data analytics, and energy management. Named among the Top 2\% Highly Influential Scientists in both 2023 and 2024 by Stanford University/Elsevier, Dr. Himeur is an Assistant Professor at the College of Engineering and Information Technology at the University of Dubai, where he also serves as Director of the Master of Electrical Engineering program. He completed his M.Sc. and Ph.D. in Electrical Engineering in 2011 and 2015, respectively, and earned the Habilitation to Direct Research in 2017, allowing him to supervise research. Dr. Himeur's academic journey includes a postdoctoral fellowship at Qatar University (2019–2022) and a tenure as Senior Researcher and Head of the TELECOM Division at the Algerian Center for Development of Advanced Technologies (CDTA). With over 180 research publications, Dr. Himeur has made valuable contributions to the field, co-leading several research projects funded by the Qatar National Research Fund (QNRF), Qatar University, and the University of Dubai. Among his numerous accolades, he was awarded the Best Research Award for the academic Year 2023/2024, the Best Student Paper Award at IEEE GPECOM 2020, and the Best Paper Award at the 11th IEEE SIGMAP in 2014. In his leadership roles, Dr. Himeur is the Publicity Chair for ICSPIS 2024, held from November 12-14, 2024, in Dubai, and the Proceedings Chair for the 10th IEEE BigDataService 2024 in Shanghai, China. He was also the Proceedings Chair for the 9th IEEE BigDataService 2023 in Athens, Greece. He serves as the Special Issue Editor for the "Internet of Energy and Artificial Intelligence for Sustainable Cities" in MDPI Energies and is the Chair of the 4th Workshop on Distributed Machine Learning for the Intelligent Computing Continuum (DML-ICC) at the IEEE/ACM UCC 2024. Additionally, Dr. Himeur chairs the Embedded Intelligence Workshop at the IEEE/ACM BDCAT conference in 2024. Beyond research, he actively teaches Advanced AI, Text Mining, and Research Methods, and supervises PhD and Master’s students. Dr. Himeur also serves as Chair of the IEEE UAE Chapter for Robotics and Autonomous Systems, Control Systems, and Engineering in Medicine and Biology, and is an Associate Editor for Frontiers in Artificial Intelligence.
\end{IEEEbiography}
\vskip 0pt plus -1fil

\begin{IEEEbiography}
[{\includegraphics[width=1in,height=1.25in,clip,keepaspectratio]{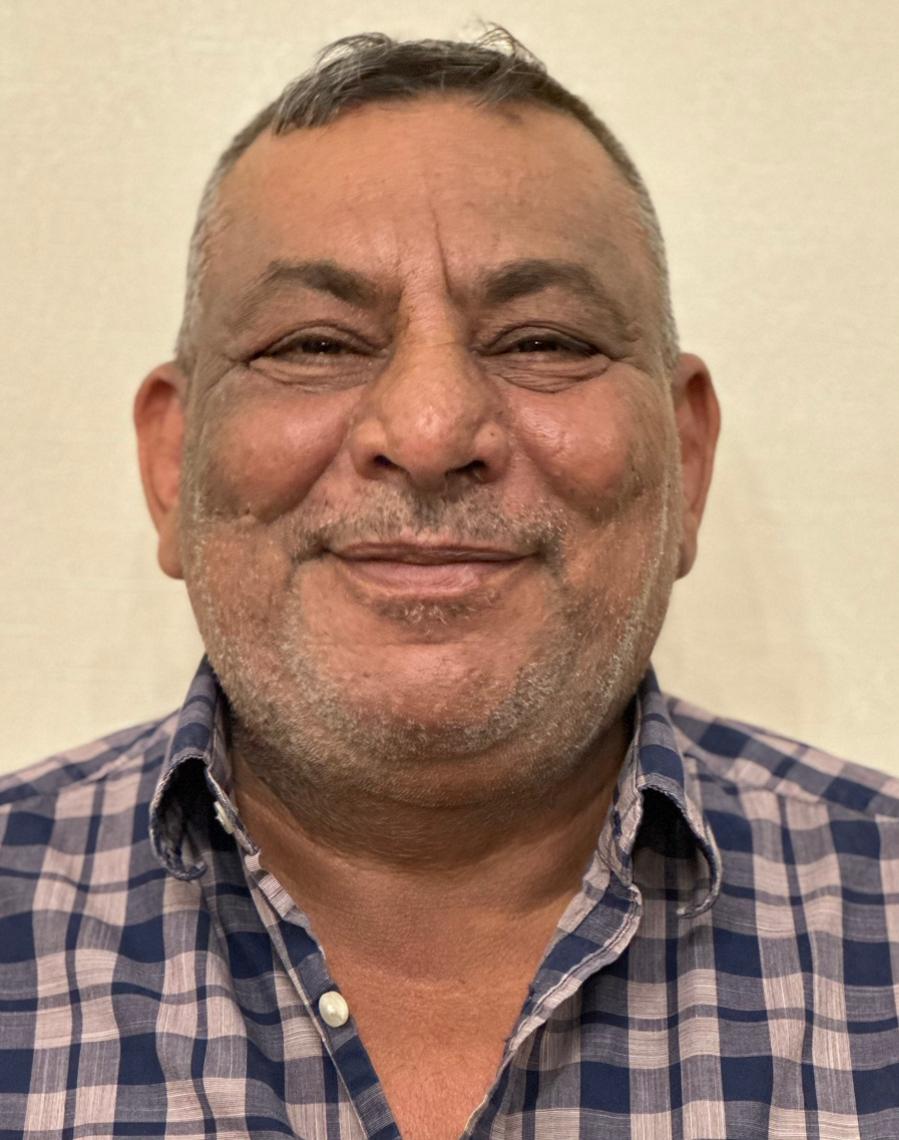}}]
{Dr. Ahmed Al Asadi} is currently a lecturer in the Department of Communication Engineering, University of Technology-Baghdad/Iraq. Dr. Asadi holds B.Sc. and M.Sc. degrees in Electronics and Communications Engineering from the University of Technology in Baghdad, Iraq (awarded in 1993 and 1996, respectively). In 2021, he further distinguished himself by completing a Ph.D. in Electrical Engineering from the University of Missouri(MU)/Columbia, USA. He has a special interest in deep learning,  machine learning. statistical signal processing and convex optimization. Dr. Al Asadi brings over 30 years of experience in teaching and researching at university.
\end{IEEEbiography}
\vskip 0pt plus -1fil

\begin{IEEEbiography}
[{\includegraphics[width=1in,height=1.25in,clip,keepaspectratio]{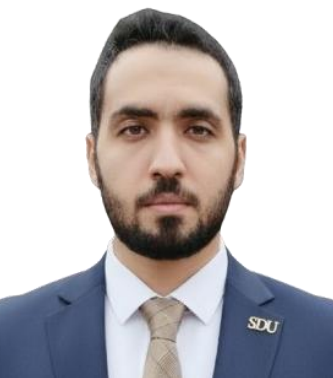}}]
{Mohammed A. Ala'anzy} (\textit{Senior Member, IEEE}) also known as Mohammed Alaa Fadhil Al-hadeethi, is an Associate Professor of Computer Science at SDU University, Kazakhstan. He earned his Ph.D. in Computer Science (Field of Study: Parallel and Distributed Computing) from Universiti Putra Malaysia (UPM) in 2023, following a Master's degree in Computer Science at UPM in 2017 and a Bachelor's degree in Computer Science from the University of Baghdad in 2014. His academic and research focus spans cloud computing, fog/edge computing, IoT, task scheduling, load balancing, energy efficiency, and algorithm design. Dr. Ala'anzy has published extensively in international journals and IEEE conferences, with contributions addressing advanced scheduling algorithms, resource allocation strategies, and smart IoT-driven applications for areas such as healthcare and Industry 4.0. He has also presented his work at numerous global conferences, including in Malaysia, Sweden, Australia, France, Romania, and the UAE, where his research has earned recognition, such as the Best Paper Award at ICoBiC 2017. Beyond his research, he is deeply engaged in academic leadership and teaching. At SDU, he lectures on Algorithms, Data Structures, Cloud Computing, Java, and Distributed Systems, and serves as a Dissertation Council Member. He actively supervises MSc and PhD students and leads funded research projects. In addition, he contributes to the wider academic community as a reviewer, program committee member, and session chair at leading international conferences. With experience in both academia and industry, including prior work as an RF Optimization Engineer with Huawei and SignalTech in Malaysia, Dr. Ala'anzy bridges theoretical and practical perspectives in computer science. His career reflects a commitment to advancing research, mentoring future scholars, and fostering innovation in emerging computing technologies.
\end{IEEEbiography}
\vskip 0pt plus -1fil

\begin{IEEEbiography}
[{\includegraphics[width=1in,height=1.25in,clip,keepaspectratio]{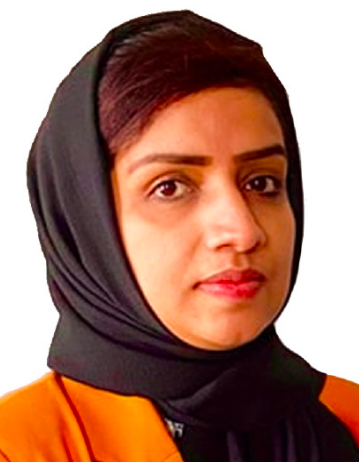}}]
{Dr. Rida Gadhafi} (Senior Member, IEEE) received the Ph.D. degree in optics and radio frequency from Université Grenoble Alpes, France, in 2013. She also held a Postdoctoral position with Technische Universität Dresden, Germany. She is currently an Assistant Professor at the College of Engineering and IT, University of Dubai. Prior to joining the University of Dubai, she worked as a Research Scientist with Khalifa University, Masdar City Campus, Abu Dhabi, since March 2015. She has published several research papers in international journals and conferences. Her main research interests include RF communication and RF (microwave/mm-wave) component design, specifically lunar communication, chipless Radio Frequency Identification tags, on-chip antennas, and PCB antennas and filters. She was a recipient of the Teaching Excellence Award from the University of Dubai in 2023, the IEEE Best Student Paper Awards in 2011 and 2012 respectively, and theIEEE MTT-S Student Sponsorship from International Microwave Symposium, Canada in 2012. She is a Vice-Chair of IEEE MTT/AP/IMS UAE chapter.
\end{IEEEbiography}
\vskip 0pt plus -1fil

\begin{IEEEbiography}
[{\includegraphics[width=1in,height=1.25in,clip,keepaspectratio]{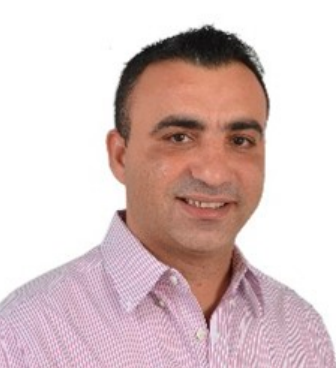}}]
{Dr. Shadi Atalla} (Member, IEEE) is an Associate Professor and Director of the Computing \& Information Systems program at the University of Dubai. With over 15 years of experience in teaching and research, he is a prominent data science evangelist and certified big data trainer, highly regarded in the industry. Dr. Atalla research focuses on developing data science algorithms, curriculum development, and artificial intelligence. He has published several papers in international scientific journals and has contributed significantly to the field of data science. Dr. Atalla also serves as the Chair of the Computer Society of IEEE UAE, showcasing his leadership qualities and dedication to advancing the field.
\end{IEEEbiography}
\vskip 0pt plus -1fil

\begin{IEEEbiography}
[{\includegraphics[width=1in,height=1.25in,clip,keepaspectratio]{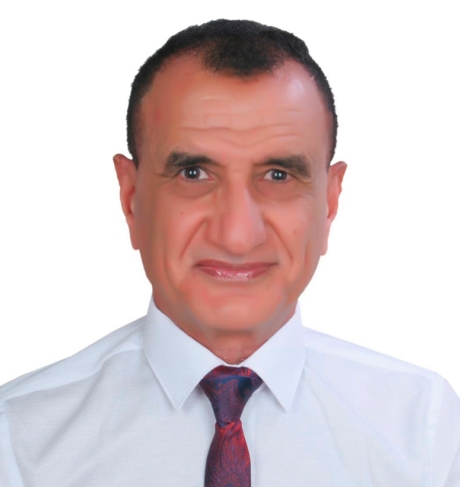}}]
{Dr. Wathiq Mansoor} (Senior Member, IEEE) is is currently the Dean of the College of Engineering at the American University in Baghdad. He has excellent academic leadership experience in well-known universities worldwide. He earned his Ph.D. in computer engineering from Aston University in UK. His doctoral work focused on the design and implementations of multiprocessor systems and communications protocols for computer vision applications. He has published many research papers in the area of Intelligent Systems, Image processing, deep learning, security, ubiquitous computing, web services, and neural networks. His current research is in the area of intelligent systems and security using neural networks with deep learning models for various applications. He has organized many international and national conferences and workshops. He is a senior member of the IEEE UAE section. He has supervised many Ph.D. and undergraduate projects in the field of Computer engineering and innovation in business, in addition to co-supervise many postgraduate students through research collaboration with
international research groups. \EOD

\end{IEEEbiography}
\vskip 0pt plus -1fil

\end{document}